\documentclass[universe,article,submit,moreauthors,pdftex]{Definitions/mdpi}

\firstpage{1} 
\pubvolume{1}
\issuenum{1}
\articlenumber{0}
\pubyear{2021}
\copyrightyear{2021}
\externaleditor{Academic Editor: { Alexei A. Starobinsky and Gerald B. Cleaver}} 
\datereceived{30 May 2021} 
\dateaccepted{30 June 2021} 
\datepublished{} 
\hreflink{https://doi.org/} 

\usepackage{bm}
\usepackage{amssymb}
\usepackage{amsmath}
\usepackage{amsfonts}
\usepackage{empheq}
\usepackage{color}

\newcommand{\be}{\begin{equation}}
\newcommand{\ee}{\end{equation}}

\newcommand{\cev}[1]{\reflectbox{\ensuremath{\vec{\reflectbox{\ensuremath{#1}}}}}}

\newcommand{\cZ}{\mathcal Z}

\newcommand{\bea}{\begin{eqnarray}}
\newcommand{\eea}{\end{eqnarray}}
\newcommand{\ben}{\begin{enumerate}}
\newcommand{\een}{\end{enumerate}}
\newcommand{\bit}{\begin{itemize}}
\newcommand{\eit}{\end{itemize}}

\def\k{\kappa}

\newcommand{\cbA}{\bm{\mathcal{A}}}
\definecolor{BrickRed}{cmyk}{0,0.89,0.94,0.28}
\definecolor{MidnightBlue}{cmyk}{0.98,0.13,0,0.43}
\definecolor{DarkGreen}{rgb}{0,0.7,0.1}

\newcommand{\bfx}{{\bf x}}
\newcommand{\bfu}{{\bf u}}
\newcommand{\bfv}{{\bf v}}

\newcommand{\bfz}{{\bf z}}
\newcommand{\bfk}{{\bf k}}
\newcommand{\bfq}{{\bf q}}
\newcommand{\bfn}{{\bf n}}
\newcommand{\bft}{{\bf t}}

\newcommand{\bfE}{{\bf E}}
\newcommand{\bfH}{{\bf H}}
\newcommand{\bfA}{{\bf A}}
\newcommand{\bfB}{{\bf B}}

\newcommand{\bfK}{{\bf K}}
\newcommand{\bfM}{{\bf M}}
\newcommand{\bfN}{{\bf N}}

\newcommand{\bfP}{{\bf P}}

\newcommand{\cF}{{\mathcal F}}

\newcommand{\cD}{{\mathcal D}}

\definecolor{BrickRed}{cmyk}{0,0.89,0.94,0.28}
\definecolor{MidnightBlue}{cmyk}{0.98,0.13,0,0.43}
\definecolor{DarkGreen}{rgb}{0,0.7,0.1}
\newcommand{\edit}[1]{{\color{black}#1}}

\usepackage{environ}
\NewEnviron{myequation}{%
\begin{equation}
\scalebox{0.88}{$\BODY$}
\end{equation}
}

\Title{Unifying Theory for Casimir Forces: Bulk and Surface Formulations}

\TitleCitation{Unifying Theory for Casimir Forces: Bulk and Surface Formulations}

\Author{{Giuseppe Bimonte} $^{1,}$*$^{,\dagger}$  and Thorsten Emig $^{\ddagger}$}

\AuthorNames{Giuseppe Bimonte,   and  Thorsten Emig}

\AuthorCitation{Bimonte, G.; Emig, T.}

\address[1]{%
 Dipartimento di Fisica E. Pancini, Universit\`{a} di
Napoli Federico II, Complesso Universitario
di Monte S. Angelo,  Via Cintia, I-80126 Napoli, Italy. \\

\corres{\hangafter=1 \hangindent=1.0em \hspace{-1em}Correspondence: giuseppe.bimonte@na.infn.it}

}

\firstnote{\hangafter=1 \hangindent=1.0em \hspace{-1em}INFN Sezione di Napoli, I-80126 Napoli, Italy. } 
\secondnote{\hangafter=1 \hangindent=1.0em \hspace{-1em}Laboratoire de Physique
Th\'eorique et Mod\`eles Statistiques, CNRS UMR 8626,
Universit\'e Paris-Saclay, \mbox{91405 Orsay CEDEX, France; thorsten.emig@universite-paris-saclay.fr.}}

\abstract{The principles of the electromagnetic fluctuation-induced phenomena such as Casimir forces are well understood. However, recent experimental advances require universal and efficient methods to compute these forces. While several approaches have been proposed in the literature, their connection is often not entirely clear, and some of them have been introduced as purely numerical techniques. Here we present a unifying approach for the Casimir force and free energy that builds on both the Maxwell stress tensor and path integral quantization. The result is presented in terms of either bulk or surface operators that describe corresponding current fluctuations. Our surface approach yields a novel formula for the Casimir free energy.
The path integral is presented both within a 
Lagrange and Hamiltonian formulation yielding different surface operators and expressions for the free energy that are equivalent. We compare our approaches to previously developed numerical methods and the scattering approach. The practical application of our methods is exemplified by the derivation of the Lifshitz formula.}


\begin{document}
 

\section{Introduction}

\begin{figure}[H]	
\includegraphics[scale=0.35]{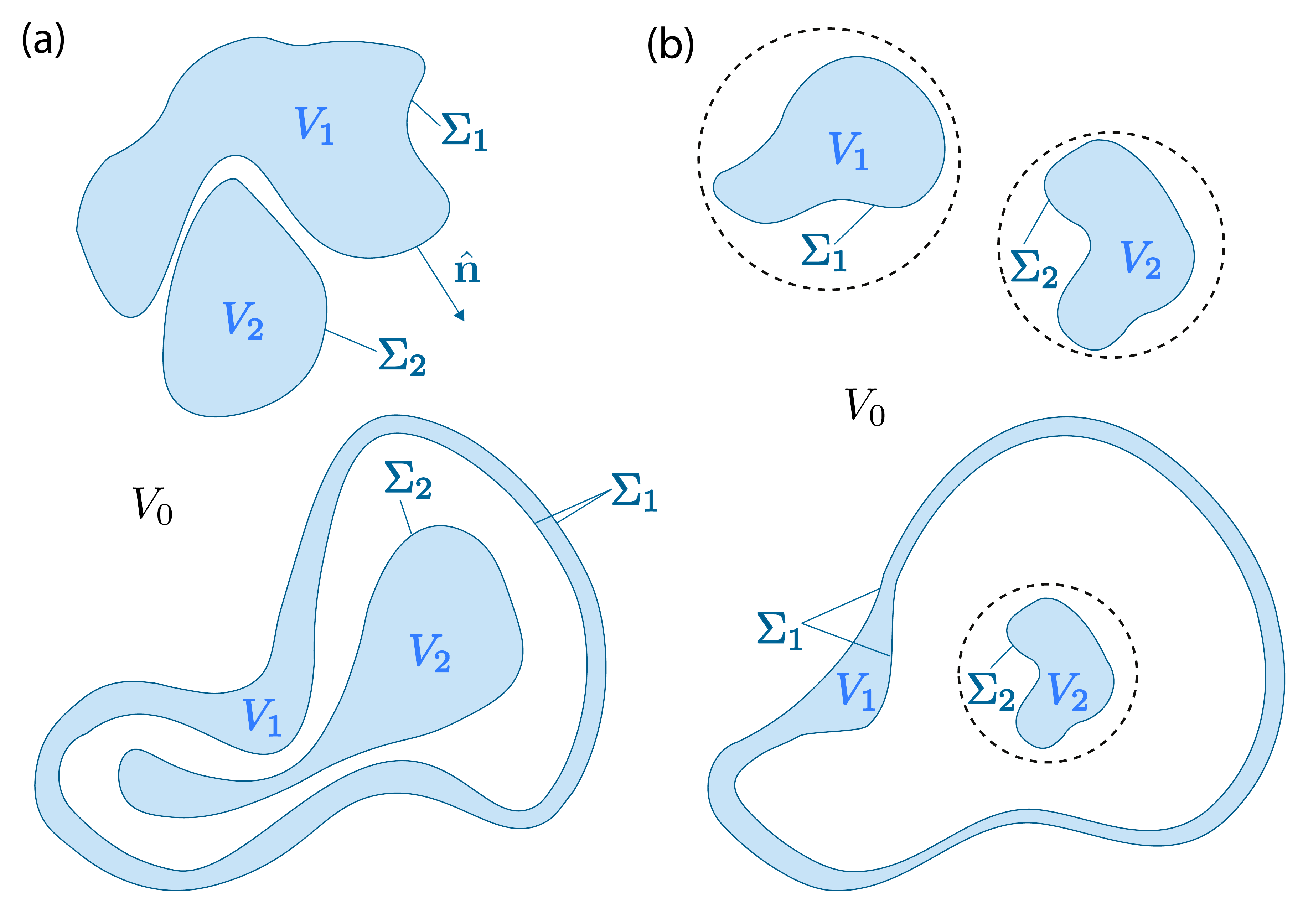}
\caption{Configuration of bodies: (\textbf{a}) general shapes and positions that can be studied with the approaches presented in this work, (\textbf{b}) non-penetrating configurations that can be studied within the scattering approach.\label{fig:configuration}}
\end{figure}

The interaction induced by quantum and thermal fluctuation of the electromagnetic field is an everyday phenomenon that acts between all neutral objects, both on atomic and macroscopic scales \cite{AnnuRev,RevModPhys2009,RevModPhys2016,Rodriguez2014,Dalvit2011}.
 For the Casimir interaction between macroscopic bodies, the last two decades have witnessed unparalleled progress in experimental observations and the development of novel theoretical approaches \cite{Rodriguez2011,Golyk}. 
In most of the recent theoretical approaches, the computation of Casimir forces between multiple objects of different shapes and material composition has been achieved by the use of scattering methods or the so-called TGTG formula \cite{Emig2007,kenneth,Neto2008,Emig2008,Rahi,Kruger2012,bimonte2009,bimonteemig,bimonte2018}. 
These approaches have the advantage of relatively low numerical effort; they are rapidly converging and can achieve in principle any desired precision \cite{Kenneth2008,Milton2008,Reid2009,Golestanian2009,Ttira2010,hartmann}. 
\edit{Another merit of these methods is the exclusion of UV divergencies by performing the subtraction analytically before any numerical computation.}
\edit{Other efficient approaches that have been developed before the scattering approaches include path integral quantizations where the boundary conditions at the surfaces are implemented by  delta functions \cite{Bordag1985}. These approaches are limited to scalar fields with Dirichlet or Neumann boundary conditions \cite{Bordag2006}, or the electromagnetic field with perfectly conducting boundary conditions \cite{Emig2003}, with the exception of a similar approach for dielectric boundaries \cite{Buscher2004}.}
However, such analytical (and semi-analytical) methods have been restricted to symmetric and simple shapes, like spheres, cylinders or ellipsoids \cite{Huth2010,Graham2011,Teo2013,Incardone2014,Emig2016}. 
\edit{Geometries where parts of the bodies interpenetrate, such as  those shown in Figure~\ref{fig:configuration}a, cannot be studied with scattering approaches.}
 For general shapes \edit{and arbitrary geometries, new methods are needed. Purely} numerical methods based on surface current fluctuations have been developed \cite{johnson}, but they rely on a full-scale numerical evaluation of matrices and their determinants, which complicates these approaches when high precision of the force is required.

Hence, there is a need to develop methods that predict Casimir interactions between objects of arbitrary geometries composed of materials with arbitrary frequency-dependent electromagnetic  properties. The Casimir force can be viewed as arising from the interaction of fluctuating currents distributions. In fact,
these effective fluctuating electric and magnetic currents can be considered to be localized either in the bulk of the bodies or just on their surfaces. The surface approach relies on the observation that the electromagnetic response of bodies can be represented entirely in terms of their surfaces, known as the “equivalence principle”, which is based on the observation that many source distributions outside a given region can produce the same field inside the region \cite{Harrington}. The surface approach has been introduced in the literature as a method for a purely { numerical} computation of Casimir interactions \cite{johnson}.  There are two different methods to implement the idea of computing Casimir forces from fluctuating currents. One can either integrate the Maxwell stress tensor over a closed surface enclosing the body, directly yielding  the Casimir force, or integrate over all electromagnetic gauge field fluctuations in a path integral, yielding the Casimir free energy. We shall consider  both approaches here. 

Compared to scattering theory-based approaches, the surface formulation has the advantage that it does not require the use of eigenfunctions of the vector wave equation that are specific to the shape of the bodies. Hence, our approach is applicable to general geometries and shapes, including interpenetrating structures. \edit{In fact, the power of the surface approach
has been demonstrated by numerical implementations in Reference \cite{johnson}, where it was used to compute the Casimir force in complicated geometries.}

In this paper, we present both the Maxwell stress tensor and path integral-based approaches for the Casimir force and free energy in terms of bulk or surface operators. Our main advancements are 
\begin{itemize}
\item A new, compact and elegant derivation of the Casimir force from the Maxwell stress tensor within both a T-operator approach and a surface operator approach;
\item \textls[-20]{A new surface formula for the Casimir free energy expressed in terms of a surface~operator;}
\item A new path integral-based derivation of a Lagrange and Hamiltonian formulation for the Casimir free energy.
\end{itemize}

We also compare the approaches presented here to  methods existing in the literature. For the special case of bodies that can be separated by non-overlapping enclosing surfaces, along which one of the coordinates in which the wave equation is separable is constant, our approach is shown to be equivalent to the scattering approach. Our approaches also show the general equivalence of the use of the Maxwell stress tensor in combination with the fluctuation-dissipation theorem on one side and the path integral representation of the Casimir force on the other side. As the most simple application of our approaches, we re-derive the Lifshitz formula for the Casimir free energy of two dielectric slabs. Other analytical applications of our approach will be presented elsewhere.

The geometries and shapes to which our approaches can be applied are shown in Figure~\ref{fig:configuration}a. For comparison, in  Figure~\ref{fig:configuration}b, we display non-penetrating bodies to which scattering theory-based approaches are limited.
In general, we assume a configuration composed of $N$ bodies with dielectric functions $\epsilon_r(\omega)$ and magnetic permeabilities $\mu_r(\omega)$, $r=1,\ldots,N$. The bodies occupy the volumes $V_r$ with surfaces $\Sigma_r$ and outward pointing surface normal vectors $\hat\bfn_r$. The space with volume $V_0$ in between the bodies is filled by matter with dielectric function $\epsilon_0(\omega)$ and magnetic permeability $\mu_0(\omega)$.

\section{Stress-Tensor Approach}

\subsection{Bulk and Surface Expressions for the Force}

Consider a collection of $N$ magneto-dielectric bodies in vacuum. In the stress-tensor approach, the (bare) Casimir force $F_i^{( {\rm bare}| r)}$ on body $r$  is obtained by integrating the expectation value $\langle T_{ij} \rangle$ of the Maxwell stress tensor  
\end{paracol}
\nointerlineskip
\be
\langle T_{ij}({\bf x})\rangle=\frac{1}{4 \pi} \left\{ \langle E_i({\bf x}) E_j({\bf x}) \rangle +\langle H_i({\bf x}) H_j({\bf x}) \rangle -\frac{1}{2} \delta_{ij}\left[ \langle E_k({\bf x}) E_k({\bf x}) \rangle + \langle H_k({\bf x}) H_k({\bf x}) \rangle \right]\right\}\label{stress}
\ee
\begin{paracol}{2}
\switchcolumn

\noindent over any closed surface ${S}_r$  drawn in the vacuum, which surrounds that body (but excludes all other bodies): 
\begin{equation}
F_i^{({\rm bare}| r)} =\oint_{{S}_r} d^2 \sigma  \,\hat { n}_j({\bf x}) \langle T_{ji} ({\bf x})\rangle \;,
\end{equation} 
where $\hat {\bf n}$ is the  unit normal oriented outside ${S}_r$, and the angular brackets denote the expectation value taken with respect to quantum and thermal fluctuations.  For a system in thermal equilibrium at temperature $T$, the  (equal-time) expectation values of the products of field components (at points ${\bf x}$ and ${\bf x}'$ in the vacuum region) are provided by the fluctuation-dissipation theorem \cite{landau,agarwal}:
\begin{eqnarray}
\langle {\hat E}_i ({\bf x})  {\hat E}_j({\bf x}') \rangle&=& 2 k_B T   \left.\sum_{n=0}^{\infty}\right.\!' \;{\cal G}^{(EE)}_{ij}({\bf x},{\bf x}';{\rm i}\,\xi_n)\;, \nonumber \\
\langle {\hat H}_i ({\bf x})  {\hat H}_j({\bf x}') \rangle&=& 2 k_B T   \left.\sum_{n=0}^{\infty}\right.\!' \;{\cal G}^{(HH)}_{ij}({\bf x},{\bf x}';{\rm i}\,\xi_n)\;,\label{correl}
\end{eqnarray}
where $\xi_n= 2 \pi n k_B T/\hbar$ are the Matsubara imaginary frequencies, and the prime in the summations mean that the $n=0$ term is taken with a weight of one half. When the r.h.s of the above equations are plugged into Equation (\ref{stress}), one obtains for $\langle T_{ij}({\bf x})\rangle$ a formally divergent expression, since the Green functions ${\cal G}^{(\alpha \beta)}_{ij}({\bf x},{\bf x}';{\rm i}\,\xi_n)$ are singular in the coincidence limit ${\bf x}={\bf x}'$. This divergence can  however be easily disposed of by noticing  that the Green's functions admit the decomposition:
\be
{\cal G}^{(\alpha \beta)}_{ij}({\bf x},{\bf x}';{\rm i}\,\xi_n)={\cal G}^{(\alpha \beta;0)}_{ij}({\bf x}-{\bf x}';{\rm i}\,\xi_n)+{\Gamma}^{(\alpha \beta)}_{ij}({\bf x},{\bf x}';{\rm i}\,\xi_n)\;,\label{splitGr}
\ee
where  ${\cal G}^{(\alpha \beta;0)}_{ij}({\bf x}-{\bf x}';{\rm i}\,\xi_n)$ is the Green's function of free space,  while ${\Gamma}^{(\alpha \beta)}_{ij}({\bf x},{\bf x}';{\rm i}\,\xi_n)$  describes  the effect of scattering of electromagnetic fields by the bodies. In the coincidence limit, only ${\cal G}^{(\alpha \beta;0)}_{ij}({\bf x}-{\bf x}';{\rm i}\,\xi_n)$ diverges, while ${\Gamma}^{(\alpha \beta)}_{ij}({\bf x},{\bf x}';{\rm i}\,\xi_n)$ attains a finite limit.  When the decomposition  in Equation (\ref{splitGr}) is used to evaluate the r.h.s. of Equation (\ref{stress}), one finds that the expectation value of the stress tensor  is decomposed in a way analogous to Equation (\ref{splitGr}):
\be
\langle T_{ij}({\bf x})\rangle= \langle T^{(0)}_{ij}({\bf x})\rangle + {\Theta}_{ij}({\bf x})\;, 
\ee  
where $\langle T^{(0)}_{ij}({\bf x})\rangle$ is the divergent expectation value of the stress tensor in empty space, and $ {\Theta}_{ij}({\bf x})$ is the  {\it finite} expression
\begin{eqnarray}
{\Theta}_{ij} ({\bf x}) &=& \frac{k_B T}{2 \pi} \left.\sum_{n=0}^{\infty}\right.\!'     \left[  {\Gamma}^{(EE)}_{ij}({\bf x},{\bf x}; {\rm i}\, \xi_n)
+ {\Gamma}^{(HH)}_{ij}({\bf x},{\bf x}; {\rm i}\, \xi_n) \right. \nonumber \\
&&\left. - \frac{1}{2} \delta_{ij} \left({\Gamma}^{(EE)}_{kk}({\bf x},{\bf x}; {\rm i}\, \xi_n)+{\Gamma}^{(HH)}_{kk}({\bf x},{\bf x}; {\rm i}\, \xi_n)\right)   \right]\;. \label{stressGamma}
\end{eqnarray}
\noindent

Since the divergent contribution $\langle T^{(0)}_{ij}({\bf x})\rangle$ is  independent of the presence of the bodies, one can  just neglect it and then one  obtains the following  finite expression for the Casimir force \edit{on body $r$ due to the presence of the other bodies},
\begin{equation}
F_i^{(r)} =\oint_{{S}_r} d^2 \sigma  \,\hat { n}_j({\bf x})\;  {\Theta}_{ji} ({\bf x}) \;.\label{force0}
\end{equation} 

The further development of the theory starts from the observation  that the  dyadic Green's functions  ${\Gamma}^{(\alpha \beta)}_{ij}({\bf x},{\bf x}')$ (for brevity, from now on we shall not display the dependence of the Green's functions on the Matsubara frequencies $\xi_n$)  can be  expressed in two distinct possible forms. 

The first representation is  general, since it  is valid  for arbitrary constitutive equations of the magneto-dielectric materials constituting the bodies, which can possibly be non-homogeneous, anisotropic, and non-local. For all points ${\bf x}$ and ${\bf x}'$,  it expresses ${\Gamma}^{(\alpha \beta)}_{ij}({\bf x},{\bf x}')$    in the form of an integral of  the  $T$-operator ${\hat T}$ (for its definition, see \mbox{Appendix \ref{app2.1}})  over the {\it volume} $V$ occupied by all bodies:  
\be
{\Gamma}^{(\alpha \beta)}_{ij}({\bf x},{\bf x}' )= \sum_{r,r'=1}^{N} \int_{V_r}  d^3 {\bf y} \int_{V_{r'}} d^3 {\bf y}'  {\cal G}^{(\alpha \rho;0)}_{ik}({\bf x}-{\bf y} )    T_{kl}^{(\rho \sigma)}({\bf y},{\bf y}')   {\cal G}_{lj}^{(\sigma \beta;0)}({\bf y}'-{\bf x}')\;.\label{repGreen1}
\ee   

The above formula has a simple intuitive interpretation, if one recalls that according to its definition,   the T-operator  provides the polarization induced in the volume of the bodies when they are  immersed in the electromagnetic field generated by a  certain distribution of external sources.
 
The second representation  is  less general than Equation (\ref{repGreen1}) because it applies only to magneto-dielectric bodies that are (piecewise) homogeneous and isotropic \footnote{This restriction is not so severe in practice, since the vast majority of Casimir experiments use test bodies that can be modelled in this way.}.     It   expresses ${\Gamma}^{(\alpha \beta)}_{ij}({\bf x},{\bf x}')$ in the form of an integral  of  the {\it surface} operator ${\hat M}^{-1}$ defined in Equation (\ref{defM}), over the union $\Sigma=\bigcup \Sigma_r$ of the surfaces $\Sigma_r$ of the bodies.  For two points ${\bf x}$ and ${\bf x}'$ both lying in the vacuum region outside the bodies\footnote{A representation analogous to Equation (\ref{repGreen2}) also exists  when one or both points belong to the regions occupied by bodies, but we shall not display it since  the surface integral expressing the force in Equation (\ref{force0}) involves only points ${\bf x}$ in the vacuum region.}, the surface representation of ${\Gamma}^{(\alpha \beta)}_{ij}({\bf x},{\bf x}')$ reads:  
\begin{eqnarray}
{\Gamma}^{(\alpha \beta)}_{ij}({\bf x},{\bf x}' )&=&- \sum_{r,r'=1}^{N} \int_{V_r}  d^3  {\bf y} \int_{V_{r'}} d^3 {\bf y}'  \,\delta(F_r ({\bf y}))\, \delta(F_{r'}({\bf y}'))\;  \nonumber \\
&\times&{\cal G}^{(\alpha \rho;0)}_{ik}({\bf x}- {\bf y}) )    \left(M^{-1}\right)_{kl}^{(\rho \sigma)}({\bf y},{\bf y}')   {\cal G}_{lj}^{(\sigma \beta;0)}({\bf y}'-{\bf x}')\;,\label{repGreen2}
\end{eqnarray}
\noindent
where $F_r({\bf y})=0$ is the equation of the surface $\Sigma_r$. 
This representation also has a simple intuitive meaning, if one considers that  $-\hat{M}^{-1}$  (see Appendix \ref{app2.2} for details) is defined  as the operator that provides the {\it fictitious} surface polarizations that radiate outside the bodies the same scattered electromagnetic field as the one radiated by the physically  induced volumic polarization, in response to an external field.  The derivations of   Equations (\ref{repGreen1}) and (\ref{repGreen2})  are presented in Appendix \ref{app2}.
It is apparent that both representations have  the same mathematical structure, consisting of a two-sided convolution of a certain kernel ${\cal K}_{ij}^{(\alpha \beta)}({\bf x},{\bf x}')$  with the free-space Green's functions $ {\cal G}_{ij}^{(\alpha \beta;0)}({\bf x}-{\bf x}')$:
\be
{\Gamma}^{(\alpha \beta)}_{ij}({\bf x},{\bf x}' )= \int_{V}  d^3{\bf y}\int_{V} d^3{\bf y}'  {\cal G}^{(\alpha \rho;0)}_{ik}({\bf x}-{\bf y} )    {\cal K}_{kl}^{(\rho \sigma)}({\bf y},{\bf y}')   {\cal G}_{lj}^{(\sigma \beta;0)}({\bf y}'-{\bf x}')\;.\label{repGreen3}
\ee 

The only difference between the two representations consists in  the expression of  ${{\cal K}}$, which in the case of  Equation (\ref{repGreen1}) is  the three-dimensional kernel $T_{kl}^{(\rho \sigma)}({\bf y},{\bf y}')$ supported in the volume $V$ occupied by the bodies:
\be
{{\cal K}}_{kl}^{(\rho \sigma)}({\bf y},{\bf y}')=T_{kl}^{(\rho \sigma)}({\bf y},{\bf y}')\;,
\ee  
while  in Equation (\ref{repGreen2}) ${{\cal K}}$ is the two-dimensional  kernel $-\left(M^{-1}\right)_{kl}^{(\rho \sigma)}({\bf y},{\bf y}')$ supported on the union $\Sigma$ of their surfaces:
\be
 {{\cal K}}_{kl}^{(\rho \sigma)}({\bf y},{\bf y}')=- \sum_{r,r'=1}^{N} \delta(F_r ({\bf y}))\, \delta(F_{r'}({\bf y}')) \left(M^{-1}\right)_{kl}^{(\rho \sigma)}({\bf y},{\bf y}') 
\ee 

In both cases,  the above equation can be concisely written  using the operator notation described in Appendix \ref{app1}:
\be
\hat{\Gamma}=\hat{\cal{G}}^{(0)}\,\hat{{\cal K}}\,\hat{\cal G}^{(0)}\;.\label{gammarepgen}
\ee
\noindent

In Appendix \ref{app3}, we prove that the structure of  the representation of ${\Gamma}^{(\alpha \beta)}_{ij}({\bf x},{\bf x}' )$ given in  Equation (\ref{repGreen3}),  allows to re-express the Casimir force Equation (\ref{force0})  in the following remarkably simple form: 
\be
{F}_i^{(r)}=2 k_B T  \left. \sum_{n=0}^{\infty}\right.\!\!'  \,   \int_{V_r}  d^3 {\bf y} \int_{V} d^3 {\bf y}'\,  \left(  {\cal K}_{l j}^{(\alpha \beta)}({\bf y}',{\bf y}; {\rm i}\, \xi_n)  \frac{\partial}{\partial {y_i}} {\cal G}^{(\beta \alpha;0)}_{jl }({\bf y}-{\bf y}';{\rm i}\, \xi_n)\right)\;.\label{force1}
\ee

A crucial role in the derivation of Equation (\ref{force1})  is  played by the fact that the kernel 
${{\cal K}}$ satisfies a set of reciprocity relations analogous to those satisfied by the Green's functions:
\be
 {\cal K}_{kl}^{(\rho \sigma)}({\bf y},{\bf y}') = (-1)^{s(\rho)+s(\sigma)}{\cal K}_{lk}^{(\sigma \rho)}({\bf y}',{\bf y}) \;,\label{recK}
\ee
where $s(E)=0$ and $s(H)=1$.  It is possible to verify that the reciprocity relations satisfied by ${\cal G}^{(\alpha \beta;0)}_{ij}({\bf y}-{\bf y}')$ and ${\cal K}^{( \alpha \beta)}_{ij}({\bf y},{\bf y}')$  ensure vanishing of the ``self-force''  ${F}_i^{({\rm self}|r)}$:
\be
{F}_i^{({\rm self}|r)}=2 k_B T  \left. \sum_{n=0}^{\infty}\right.\!\!'  \,   \int_{V_r}  d^3{\bf y} \int_{V_r} d^3 {\bf y}'\,  \left(  {\cal K}_{ij}^{(\alpha \beta)}({\bf y}',{\bf y}; {\rm i}\, \xi_n)  \frac{\partial}{\partial {y_i}} {\cal G}^{(\beta \alpha;0)}_{ji}({\bf y}-{\bf y}';{\rm i}\, \xi_n)\right)=0\;.
\ee

This implies that in Equation (\ref{force1})  the ${\bf y}'$ integral is in fact restricted to ${V}-{V}_r$, in accord with one's intuition that the force on body $r$ is due to the interaction with the other bodies. It is possible to present Equation (\ref{force1}) in a more compact and symmetric form, by defining the derivative    $\partial/\partial {\bf x}_r$ of any kernel  ${\cal A}({\bf y},{\bf y}')$, with respect to rigid translations of the r-th~body:
\be
\frac{\partial}{\partial {\bf x}_r} {\cal A}({\bf y},{\bf y}') \equiv \psi_r({\bf y})\frac{\partial}{\partial {\bf y}}{\cal A}({\bf y},{\bf y}') +\psi_r({\bf y}')\frac{\partial}{\partial {\bf y}'}{\cal A}({\bf y},{\bf y}')\;,
\ee
where $\psi_r({\bf y})$ is the characteristic function of $V_r$: $\psi_r({\bf y})=1$ if ${\bf y} \in V_r$,  $\psi_r({\bf y})=0$ if ${\bf y} \notin V_r$. Using $\partial/\partial {\bf x}_r$, we can rewrite Equation (\ref{force1}) as
\be
{\bf F}^{(r)}=k_B T  \left.\sum_{n=0}^{\infty}\right.\!\!'  \,   \int_{V}  d^3{\bf y}\int_{V} d^3{\bf y}' \left(  {\cal K}_{ij}^{(\alpha \beta)}({\bf y}',{\bf y}; {\rm i}\, \xi_n)  \frac{\partial}{\partial {\bf x}_r} {\cal G}^{(\beta \alpha;0)}_{ji}({\bf y}-{\bf y}';{\rm i}\, \xi_n)\right) \;.\label{force}
\ee

The  expression on the r.h.s. of the above formula can  be compactly expressed using the operator notation and the trace operation described in Appendix \ref{app1}:
\be
{\bf F}^{(r)}=k_B T  \left.\sum_{n=0}^{\infty}\right.\!\!'  \,{\rm Tr} \left[ {\hat {\cal K}}({\rm i}\, \xi_n) \frac{\partial}{\partial {\bf x}_r} \hat{\cal G}^{(0)} ({\rm i}\, \xi_n)\right]\;.\label{force}
\ee

Depending on whether we  use for the kernel ${\hat K}$ the $T$-operator of Equation (\ref{repGreen1}) or rather the surface operator $-{\hat M}$ of Equation (\ref{repGreen2}),  Equation (\ref{force}) provides us with  two distinct but formally similar representations of the Casimir force,  which is expressed either as an integral over the volume $V$ occupied by the bodies or as an integral over their  surfaces $\Sigma$. One feature of Equation (\ref{force}) is worth stressing. Since  the force is expressed as a trace,   Equation (\ref{force}) can be evaluated in an {\it arbitrary} basis, leaving one with complete freedom in the choice of the most convenient basis in a concrete situation. 
A representation of the Casimir force in the form of a volume integral  equivalent to Equation (\ref{force}) was derived in~\cite{Kruger2012}, while the surface-integral  representation was obtained in \cite{johnson}.    Equation (\ref{force}) can be computed numerically for any shapes and dispositions of the bodies, by using discrete meshes covering the bodies. An efficient numerical scheme based on surface-elements methods is described in  \cite{johnson}, where it was used to compute the Casimir force in complex geometries, not amenable to analytical techniques. 

\subsection{Casimir Free Energy}

\textls[-15]{In this section, we compute the Casimir free energy ${\cal F}$ of  the system of bodies, starting from the force formula Equation (\ref{force}). We shall see that the T-operator and the surface-operator approaches lead to two distinct but  equivalent representations of the Casimir~energy.}

\subsubsection{T-Operator Approach}

Plugging into Equation (\ref{force}) the expression of the T-operator Equation (\ref{Top}), we find that the Casimir force can be expressed in the form:
\be
{\bf F}^{(r)}= k_B T \left. \sum_{n=0}^{\infty}\right.\!\!' \, {\rm Tr}\left( {\hat T}({\rm i}\, \xi_n) \frac{\partial}{\partial {\bf x}_r} \hat{\cal G}^{(0)}({\rm i}\, \xi_n)\right) =k_B T \left. \sum_{n=0}^{\infty}\right.\!\!' \, {\rm Tr}\left[ \frac{1}{1-{\hat \chi}\, \hat{\cal G}^{(0)}}\, \frac{\partial}{\partial {\bf x}_r} \left( {\hat{\chi}} \,\hat{\cal G}^{(0)}\right)\right]\;,
\ee
where in the last passage, we made use of the fact that the polarization operator ${\hat \chi}$ defined in Equation (\ref{defchi}) is invariant under a rigid displacement of the body. The r.h.s. of the above equation can be formally expressed as a gradient:
\be
F^{(r)}=- \frac{\partial}{\partial {\bf x}_r}\,{\cal F}_{\rm bare}\;,
\ee  
where ${\cal F}_{\rm bare}$ is the  {\it bare} free energy:
\be
{\cal F}_{\rm bare}= k_B T \left. \sum_{n=0}^{\infty}\right.\!\!'\,  {\rm Tr}\; \log [1-{\hat{\chi}} \,\hat{\cal G}^{(0)}]\; 
\ee

Unfortunately, ${\cal F}_{\rm bare}$ is formally divergent. To obtain the finite Casimir free energy ${\cal F}$, one has to subtract  from ${\cal F}_{\rm bare}$  the divergent self-energies ${\cal F}^{(r)}_{\rm self}$ of the individual bodies:
\be
{\cal F}^{(r)}_{\rm self}=   \left\{k_B T \left. \sum_{n=0}^{\infty}\right.\! \!' \, {\rm Tr}\; \log [1-{\hat{\chi}_r} \,\hat{\cal G}^{(0)}] \right\}\;,
\ee
where $\hat{\chi}_r = \hat{\psi}_r \hat{\chi} \hat{\psi}_r$ is the polarizability operator of body $r$ in isolation.
In the   case of a system composed by two bodies,  the renormalized Casimir free energy  can be recast in the following TGTG form:
\be
{\cal F}={\cal F}_{\rm bare}- {\cal F}^{(1)}_{\rm self}-{\cal F}^{(2)}_{\rm self}=k_B T \left. \sum_{n=0}^{\infty}\right.\!\!' \, {\rm Tr}\; \log [1-{\hat{T}_1} \,\hat{\cal G}^{(0)} \,{\hat{T}_2} \,\hat{\cal G}^{(0)}]\,,\label{TGTG}
\ee
where
\be
{\hat T}_r=\frac{1}{1-{\hat \chi}_r\, \hat{\cal G}^{(0)}}\,{\hat \chi}_r\,,
\ee
is  the T-operator of body $r$ in isolation.  To prove Equation (\ref{TGTG}), one notes that for each Matsubara mode the operator identity holds:
\be
(1-\hat{\cal G}^{(0)} \hat{\chi}_1)\hat{\cal G}^{(0)} \hat{T}_1 
\hat{\cal G}^{(0)} \hat{T}_2 
(1-\hat{\cal G}^{(0)} \hat{\chi}_2)
=\hat{\cal G}^{(0)} \hat{\chi}_1\hat{\cal G}^{(0)}\hat{\chi}_2\;.
\ee

The above identity in turn allows to prove the following chain of identities:
$$
{\rm Tr} \log[1-\hat{\cal G}^{(0)} \hat{\chi}_1]+{\rm Tr}\log [1- \hat{\cal G}^{(0)} \,{\hat{T}_1} \,\hat{\cal G}^{(0)}{\hat{T}_2}]+{\rm Tr} \log[1-\hat{\cal G}^{(0)} \hat{\chi}_2]
$$
$$
={\rm Tr} \log[(1-\hat{\cal G}^{(0)} \hat{\chi}_1)(1- \hat{\cal G}^{(0)} \,{\hat{T}_1} \,\hat{\cal G}^{(0)}{\hat{T}_2})(1-\hat{\cal G}^{(0)} \hat{\chi}_2)]
$$
$$
={\rm Tr} \log[(1-\hat{\cal G}^{(0)} \hat{\chi}_1)(1-\hat{\cal G}^{(0)} \hat{\chi}_2)-(1-\hat{\cal G}^{(0)} \hat{\chi}_1)\hat{\cal G}^{(0)} \,{\hat{T}_1} \,\hat{\cal G}^{(0)}{\hat{T}_2}(1-\hat{\cal G}^{(0)} \hat{\chi}_2)]
$$
$$
={\rm Tr} \log[(1-\hat{\cal G}^{(0)} \hat{\chi}_1)(1-\hat{\cal G}^{(0)} \hat{\chi}_2)-\hat{\cal G}^{(0)} \hat{\chi}_1\hat{\cal G}^{(0)}\hat{\chi}_2 ]={\rm Tr} \log[(1-\hat{\cal G}^{(0)}( \hat{\chi}_1+\hat{\chi}_2)]
$$
\be
={\rm Tr} \log[1-\hat{\cal G}^{(0)}\hat{\chi}]\;.
\ee

Equating the first line with the last line, we obtain the identity:
\begin{eqnarray}
 &&{\rm Tr} \log[1-\hat{\cal G}^{(0)}\hat{\chi}] \nonumber\\
&&={\rm Tr} \log[1-\hat{\cal G}^{(0)} \hat{\chi}_1]+{\rm Tr}\log [1- \hat{\cal G}^{(0)} \,{\hat{T}_1} \,\hat{\cal G}^{(0)}{\hat{T}_2}]+{\rm Tr} \log[1-\hat{\cal G}^{(0)} \hat{\chi}_2]\;.
\end{eqnarray}
\noindent

Upon summing the above identity  over all Matsubara modes (with weight one half for the $n=0$ term), and then multiplying it by $k_B T$, we find:
\be
{\cal F}^{(1)}_{\rm self}+{\cal F}^{(2)}_{\rm self}+{\cal F}={\cal F}_{\rm bare}\;,
\ee
which is equivalent to Equation (\ref{TGTG}). The energy formula Equation (\ref{TGTG}) was derived in~\cite{kenneth} using the path-integral method and in \cite{Kruger2012}, using Rytov's  fluctuational electrodynamics~\cite{Rytov}. 

\subsubsection{Surface Operator Approach}

Now we derive the surface-operator representation of the Casimir energy. To do that, we start from the surface-operator representation of the force, which  is obtained by replacing  $\hat{K}$ in Equation (\ref{force}) with minus the inverse of the surface operator $\hat{M}$ defined in Equation (\ref{defM}):
\be
{\bf F}^{(r)}=-k_B T  \left.\sum_{n=0}^{\infty}\right.\!\!'  \,{\rm Tr} \left[ {\hat {M}^{-1}}({\rm i}\, \xi_n) \frac{\partial}{\partial {\bf x}_r} \hat{\cal G}^{(0)} ({\rm i}\, \xi_n)\right]\;.
\ee

This can also be  written as:
\be
{\bf F}^{(r)}=-k_B T  \left.\sum_{n=0}^{\infty}\right.\!\!'  \,{\rm Tr} \left[ {\hat {M}^{-1}}({\rm i}\, \xi_n) \frac{\partial}{\partial {\bf x}_r} \left(\hat{\Pi}\hat{\cal G}^{(0)} ({\rm i}\, \xi_n) \hat{\Pi} \right)\right]\;, \label{surffo0}
\ee
where $\hat{\Pi}$ is the tangential projection operator defined in Appendix \ref{app2.2}. Now, one notes the~identity:
\be 
  \frac{\partial}{\partial {\bf x}_r} \left(\hat{\Pi}\,\hat{\cal G}^{(0)} ({\rm i}\, \xi_n)\, \hat{\Pi} \right)= \frac{\partial  {\hat M} }{\partial {\bf x}_r}\;,\label{ident0}
\ee
which  is a direct consequence of Equation (\ref{defM}) since 
\be
\frac{\partial }{\partial {\bf x}_r}\sum_{s=1}^N\hat{\Pi}_s \,   \hat{\cal G}^{(s)}\,\hat{\Pi}_s=0\;.
\ee

Plugging Equation (\ref{ident0}) into Equation (\ref{surffo0}), we obtain:
\be
F^{(r)}= -k_B T \left. \sum_{n=0}^{\infty}\right.\!\!' \, {\rm Tr} \left[ {\hat M}^{-1}({\rm i}\, \xi_n) \frac{\partial}{\partial {\bf x}_r} \hat{M}({\rm i}\, \xi_n)\right] \;,
\ee

The r.h.s. of the above equation can be formally expressed as a gradient:
\be
F^{(r)}=- \frac{\partial}{\partial {\bf x}_r}\,\tilde{{\cal F}}_{\rm bare}\;,
\ee 
where $\tilde{{\cal F}}_{\rm bare}$ is the  {\it bare} free energy:
\be
{\tilde {\cal F}}_{\rm bare}= k_B T \left. \sum_{n=0}^{\infty}\right.\!\!'\,  {\rm Tr}\; \log  {\hat M}({\rm i}\, \xi_n)\;. \label{bareensurf}
\ee 

Similarly to what we found in the T-operator approach, the surface formula of the bare-energy ${\tilde {\cal F}}_{\rm bare}$ is  formally divergent. The finite Casimir free energy  is obtained by subtracting from  ${\tilde {\cal F}}_{\rm bare}$  the limit  $\tilde{{\cal F}}^{(\infty)}_{\rm bare}$ of the bare energy when the bodies are taken infinitely apart from each other. From Equation (\ref{defM}), one sees that in the limit of infinite separations, the operator  ${\hat M}$ approaches the limit ${\hat M}_{\infty}$
\be
{\hat M}_{\infty}=   \sum_{r=1}^N   {\hat M}_r\;,\label{defMinf}
\ee
where
\be
{\hat M}_{r}=  \hat{\Pi}_r\,(  \hat{\cal G}^{(r)}+  \hat{\cal G}^{(0)})\,\hat{\Pi}_r  \;.\label{defMbodies}
\ee

Notice that the surface operator ${\hat M}_{r}$ is localized onto the surface $\Sigma_r$ of the $r$-th body. This implies that:
\be
{\hat M}_{r}\, {\hat M}_{s}=0\;,\;\;\;\;\;{\rm for}\;\;r \neq s\;.
\ee

Using Equation (\ref{defMinf}), we find that    $\tilde{{\cal F}}^{(\infty)}_{\rm bare}$ is the formally divergent quantity:
\be
\tilde{{\cal F}}^{(\infty)}_{\rm bare}= k_B T \left. \sum_{n=0}^{\infty}\right.\!\!'\, \; {\rm Tr}\; \log  {\hat M}_{\infty}({\rm i}\, \xi_n)=\sum_{r=1}^{N} k_B T \left. \sum_{n=0}^{\infty}\right.\!\!'\, \;  {\rm Tr}\; \log  {\hat M}_{r}({\rm i}\, \xi_n)\equiv \sum_{r=1}^{N} \tilde{{\cal F}}^{(r)}_{\rm self}\;.
\label{selfensurf}
\ee

The additive character of $\tilde{{\cal F}}^{(\infty)}_{\rm bare}$ allows to interpret $ \tilde{{\cal F}}^{(r)}_{\rm self}$ as representing the (infinite) the self-energy of the bodies in the surface approach.
Upon subtracting Equation (\ref{selfensurf}) from Equation (\ref{bareensurf}), we arrive at  the following formula for the Casimir energy:
\be
{\cal F}=k_B T \left. \sum_{n=0}^{\infty}\right.\!\!'\,  \log \det \frac{ {\hat M}({\rm i}\, \xi_n)}{{\hat M}_{\infty}({\rm i}\, \xi_n)}\;. \label{ensurf}
\ee

An easy computation shows that 
\be
 \frac{1}{{\hat M}_{\infty}}\,   {\hat M}=1+\sum_{r \neq s}\frac{1}{\hat{M}_r} \hat{\cal G}_{rs}^{(0)}\;.
\ee
where
\be
\hat{\cal G}_{rs}^{(0)}= \hat{\Pi}_r \,\hat{\cal G}^{(0)}\, \hat{\Pi}_s\;.
\ee

Substitution of the above formula into Equation (\ref{ensurf}) results in the following surface formula for the Casimir energy:
\be
{\cal F}=k_B T \left. \sum_{n=0}^{\infty}\right.\!' \log \det \left[1+\sum_{r \neq s}\frac{1}{\hat{M}_r} \hat{\cal G}_{rs}^{(0)}  \right] \;.\label{ensurF}
\ee

In the simple case of two bodies, the above formula reduces to:
\be
{\cal F}=k_B T \left. \sum_{n=0}^{\infty}\right.\!'  {\rm Tr}\,\log \left[1- \frac{1}{\,{\hat M}_1} \, \hat{\cal G}_{12}^{(0)} \frac{1}{\,{\hat M}_2}  \hat{\cal G}_{21}^{(0)}  \right]\;.\label{renensurf}
\ee

The surface formulas for the Casimir energy given  in Equations (\ref{ensurF}) and (\ref{renensurf}) were not known before and are presented here for the first time.
Comparison of Equation (\ref{renensurf}) with Equation (\ref{TGTG}) reveals the striking similarity of the T-operator and surface-approach representations of the Casimir energy. Indeed we see that both formulas can be written in the form:
\be
{\cal F}=k_B T \left. \sum_{n=0}^{\infty}\right.\!'  {\rm Tr}\,\log \left[1- \hat{\cal{K}}_1 \, \hat{\cal G}^{(0)} \hat{\cal{K}}_2 \hat{\cal G} ^{(0)} \right]\;,\label{renensurfgen}
\ee
where $\hat{\cal{K}}_r$ is   the   kernel, which gives the scattering Green's function of body $r$ in isolation:
\be
\hat{\Gamma}_r=\hat{\cal G}^{(0)}\,\hat{{\cal K}}_r\,\hat{\cal G}^{(0)}\;. \label{gammarepgenis}
\ee

\section{Equivalence of the Surface-Formula with the Scattering Formula for the Casimir~Energy}

In the previous sections, we have shown that, both in the $T$-operator   and in the surface approaches, the Casimir  energy ${\cal F}$ of two bodies can be expressed by the general \mbox{Equation (\ref{renensurfgen})}.
This formula is valid {for any shape and relative dispositions} of the two bodies, and  in particular  for two  interleaved  bodies (see Figure~\ref{fig:configuration}a).  Now we show that when the two bodies can be enclosed within two non-overlapping spheres (see Figure~\ref{fig:configuration}b),  \mbox{Equation (\ref{renensurfgen})}   is  the same as the well-known  { scattering} formula \cite{Emig2007,Neto2008,Emig2008,Rahi,Kruger2012}:
\be
{\cal F}=k_B T \left. \sum_{n=0}^{\infty}\right.\!' \rm{tr}  \log \left[1- {\cal T}^{(1)} \,  {\cal U}^{(12)}  {\cal T}^{(2)} \,  {\cal U}^{(21)} \right] \; .\label{scaterfor}
\ee  
where ${\cal T}^{(r)}$ is the {scattering matrix} of  body $r$ (see Equation (\ref{tmatdef}) for the definition of ${\cal T}^{(r)}$),  ${\cal U}^{(rs)}$ are the translation matrices defined in Equation (\ref{transl}) and $\rm{tr}$ denotes a trace over multipole indices.
 
\noindent

To prove equivalence of Equation (\ref{renensurfgen}) with Equation (\ref{scaterfor}), one starts from the observation that the trace operation in Equation (\ref{renensurfgen}) involves evaluating  the  Green functions ${\stackrel{\leftrightarrow}{\cal G}}^{(\alpha \beta;0)}\!\!\!\!({\bf y},{\bf y}')$   at points ${\bf y}$ and ${\bf y}'$,  one of which (call it ${\bf y}_1$) belongs to  body 1, while the other (call it ${\bf y}_2$) belongs to body  2. For two bodies that can be separated by non-overlapping spheres, it is warranted that $|{\bf y}_1-{\bf X}_1| < |{\bf y}_2-{\bf X}_1|$ and 
$|{\bf y}_2-{\bf X}_2| < |{\bf y}_1-{\bf X}_2|$,  where  ${\bf X}_1$ and ${\bf X}_2$ are the positions of the centers of the spheres $S^{(1)}$ and $S^{(2)}$, respectively, and $d=|{\bf X}_2-{\bf X}_1|$ is their distance. This condition satisfied by ${\bf y}_1$ and ${\bf y}_2$ ensures that it is legitimate to express ${\stackrel{\leftrightarrow}{\cal G}}^{(\alpha \beta;0)}\!\!\!\!({\bf y}_r,{\bf y}_s)$ (with $r\neq s=1,2$) by the  partial-wave expansion (see Appendix \ref{app4}):
\begin{eqnarray}
{\stackrel{\leftrightarrow} { {\cal G}}}^{(\alpha \beta;0)} \!\!\!\!\!  \!\!\!\!\! ({\bf y}_r,{\bf y}_s)\!\!\!\!&=&\!\!\! \!\lambda (-1)^{s(\beta)}\,\sum_{plm}   { \Phi}^{(\alpha | {\rm reg})}_{plm}({\bf y}_r-{\bf X}_r) \otimes { \Phi}^{(\beta | {\rm out})}_{pl-m}({\bf y}_s-{\bf X}_r) \nonumber \\
&=&\!\!\!\lambda\, (-1)^{s(\beta)}\sum_{plm}  \sum_{p'l'} { \Phi}^{(\alpha | {\rm reg})}_{p lm}({\bf y}_r-{\bf X}_r) \otimes  {\cal U}^{(rs)}_{p l ; p' l'}(d)\, { \Phi}^{(\beta | {\rm reg})}_{p'l'-m}({\bf y}_s-{\bf X}_s)\;,
\end{eqnarray}
where ${ \Phi}^{ {\rm( reg/out})}_{p lm}({\bf y}_r-{\bf X}_r)$ are a basis of regular and outgoing spherical waves with origin at ${\bf X}_r$.
When the above   expansion is substituted into Equation (\ref{renensurfgen}) and the trace is evaluated,  one finds that ${\cal F}$ can be recast in the form:
\be
{\cal F}=k_B T \left. \sum_{n=0}^{\infty}\right.\!'  \;\rm{tr} \;\log \left[1-{\cal N} \right]\;.\label{firststep}
\ee
where ${\cal N}$ is the matrix of elements:
\end{paracol}
\nointerlineskip
\begin{eqnarray}
&&{\cal N}_{plm;p' l' m'} \equiv \;    \sum_{p'' l''}  \sum_{p''' l''' m'''}  \sum_{p'''' l''''} \; {\cal U}^{(21)}_{p l,p''l''} (d) \; {\cal U}^{(12)}_{p''' l''' ; p'''' l''''}(d) \nonumber \\
&\times& \lambda   \sum_{\alpha, \mu} (-1)^{s(\alpha)}  \int_{{V}_1} d^3 {\bf y}_1  \int_{{V}_1} d^3 {\bf y}'_1  \; { \Phi}^{(\alpha | {\rm reg})}_{p'' l''-m}({\bf y}_1-{\bf X}_1) \cdot {\stackrel{\leftrightarrow} {\cal K}}^{(\alpha \mu)}_1 \;({\bf y}_1,{\bf y}'_1)\cdot { \Phi}^{(\mu | {\rm reg})}_{p''' l'''m'''}({\bf y}'_1-{\bf X}_1) \; \nonumber \\ 
&\times& \lambda    \sum_{\beta,\nu} (-1)^{s(\nu)}  \int_{{V}_2} d^3 {\bf y}_2  \int_{{V}_2} d^3 {\bf y}'_2\;{ \Phi}^{(\nu | {\rm reg})}_{p''''l''''-m'''}({\bf y}_2-{\bf X}_2)\, \cdot {\stackrel{\leftrightarrow}{\cal K} }^{(\nu \beta)}_2({\bf y}_2,{\bf y}'_2) \cdot { \Phi}^{(\beta | {\rm reg})}_{p' l' m'}({\bf y}'_2 -{\bf X}_2)\;. \nonumber
\end{eqnarray}
\begin{paracol}{2}
\switchcolumn
\noindent

Recalling the formula Equation (\ref{tmatK}) for the scattering matrices ${\cal T}^{(r)}$ of the two bodies, we  see that ${\cal N}$ is the matrix:  
\be
{\cal N} ={\cal U}^{(21)} (d) \;{\cal T}^{(1)} {\cal U}^{(12)}(d)\,{\cal T}^{(2)}\;.
\ee

Upon substituting  the above expression into the r.h.s. of Equation (\ref{firststep}), and using cyclicity of the trace,  we see that   Equation (\ref{firststep})  indeed coincides with the scattering formula Equation (\ref{scaterfor}).

\section{Path Integral Approach}

As in the previous sections, we consider again $N$ dielectric bodies occupying the volumes $V_r$, $r=1,\ldots, N$, bounded by surfaces $\Sigma_r$. Their electromagnetic  properties are described by the dielectric functions $\epsilon^{(r)}$ and magnetic permeability $\mu^{(r)}$. The bodies are embedded in a homogeneous medium occupying the outside volume of the bodies, $V_0$, with dielectric function $\epsilon^{(0)}$ and magnetic permeability $\mu^{(0)}$.

In the Euclidean path integral quantization of the electromagnetic field, the Casimir free energy at finite temperature $T$ can be obtained as
\be
\label{PI_free_energy}
\cF = - k_B T \left. \sum_{n=0}^{\infty}\right.\!'
\log\frac{\cZ(\kappa_n)}{\cZ_{\infty}(\kappa_n)} \, ,
\ee 
where the sum runs over the Matsubara momenta $\kappa_n=2\pi n k_B T/\hbar c$, with a weight of $1/2$ for $n=0$. The partition function $\cZ$ is given by a path integral that we shall derive now. The partition function $\cZ_{\infty}$ describes the configuration of infinitely separated bodies and subtracts the self-energies of the bodies from the bare free energy. In the following two sections, we shall derive both a Lagrangian and a Hamiltonian path integral expression of the partition function. In both cases, we employ a fluctuating { surface} current approach. A path integral approach that is based on bulk currents can be found, e.g., in Reference~\cite{Rahi}.

\subsection{Lagrange Formulation}

The action of the electromagnetic field coupled to bound sources $\bfP_\text{ind}$ in the absence of free sources is in general given by
\begin{equation}
  \label{eq:2}
  S_\text{EM} = \int d^3 {\bf x}\, \left[  \frac{1}{2} \left(
      \epsilon_\bfx \bfE^2 -  \frac{1}{\mu_\bfx} \bfB^2 \right)
     + \bfP_\text{ind} \cdot \bfE \right] \, .
\end{equation}

In the following, we express the action in terms of the gauge field $\bfA$ choosing the transverse or temporal gauge with $A_0=0$. The functional integral will then run over $\bfA$ only. The electric field is given by $\bfE = i k \bfA \to -\k \bfA$ and the magnetic field by $\bfB = \nabla \times \bfA$. Then the action  in terms of the induced sources at fixed frequency $\k$ is given by
\begin{eqnarray}
\label{eq:EM_hatS}
  \hat S[\bfA] &=& -\frac{1}{2} \int_{\mathbb{R}^3} d^3 {\bf x}\, \left[
                        \bfA^2 \epsilon_\bfx \k^2 + \frac{1}{\mu_\bfx} (\nabla \times \bfA)^2\right]
- \k \sum_{r=1}^N \int_{V_r} d^3 {\bf x} \,\bfA \cdot \bfP_r \, ,
\end{eqnarray}
for fluctuations $\bfA$ of the gauge field, and induced bulk
currents $\bfP_r$ inside the objects. The inverse of the kernel
of the quadratic part of this action is given by the Green tensor
$G(\bfx,\bfx')$, which is defined by
\begin{equation}
  \label{eq:G_dyadic_def}
\nabla \times \frac{1}{\mu_\bfx} \nabla  \times  {\stackrel{\leftrightarrow}{\cal G}}^{(AA)}(\bfx,\bfx') +
\epsilon_\bfx \k^2 \, {\stackrel{\leftrightarrow}{\cal G}}^{(AA)}(\bfx,\bfx') = 4\pi\, {\bf 1} \, \delta(\bfx-\bfx') \, .
\end{equation}

For spatially constant $\epsilon$ and $\mu$ with body $r$, this yields the free Green's tensor
\begin{equation}
  \label{eq:G_dyadic}
  {\stackrel{\leftrightarrow}{\cal G}}^{(AA;r)}(\bfx,\bfx') = \mu_r\left( {\bf 1}
    -\frac{1}{\epsilon_r \mu_r\k^2} \nabla\otimes\nabla \right)
  \frac{e^{-\sqrt{\epsilon_r \mu_r}\k |\bfx-\bfx'|}}{
    |\bfx-\bfx'|} \, ,
\end{equation}
which is symmetric, reflecting reciprocity. From the relation between the gauge field $\bfA$ and the electric field $\bfE$ follows the relation $-\k^2{\stackrel{\leftrightarrow}{\cal G}}^{(AA;r)}(\bfx,\bfx')={\stackrel{\leftrightarrow}{\cal G}}^{(EE;r)}(\bfx,\bfx') $, which allows to compare the results below to those of the stress-tensor-based derivation.

Next, we define the classical solutions
$\cbA_r$ of the vector wave equation in each region 
$V_r$, obeying
\begin{equation}
  \label{eq:WE_E}
  \nabla \times \nabla \times \cbA_r + \epsilon_r \mu_r
  \k^2 \cbA_r  = - \k \mu_r \bfP_r 
\end{equation}

We use this definition together with the fact that $\bfA$ has no
sources inside $V_r$, i.e., obeys above wave equation with
vanishing right-hand side, to rewrite the source terms of Equation~(\ref{eq:EM_hatS}) as
\begin{eqnarray}
  \label{eq:EM_source_term}
  -\k \int_{V_r} d^3 {\bf x} \,\bfA \cdot \bfP_r &=&
                \int_{V_r} d^3 {\bf x}      \,\bfA \cdot \left[
 \frac{1}{\mu_r}\nabla\times\nabla \times
                                                             \cbA_r+
                                                             \epsilon_r
                                                             \k^2
                                                             \cbA_r
                                                             \right]
  \\
  &=& \frac{1}{\mu_r} \int_{V_r} d^3 {\bf x} \,     \,\left[ \bfA \cdot
      ( \nabla\times\nabla\times \cbA_r)  -  (
      \nabla\times\nabla\times \bfA)  \cdot \cbA_r \right]\nonumber\\
  &=&  \frac{1}{\mu_r} \int_{V_r} d^3 {\bf x} \, \left[ \nabla \cdot 
      ((\nabla\times \cbA_r)\times\bfA )-\nabla \cdot ((\nabla\times
      \bfA)\times\cbA_r)\right]\nonumber\\
  &=&  \frac{1}{\mu_r} \int_{\Sigma_r} d^3 {\bf x} \, \left[ \bfn_r \cdot 
      ((\nabla\times \cbA_r)\times\bfA )-\bfn_r \cdot ((\nabla\times
      \bfA)\times\cbA_r)\right]\nonumber\\
  &=& \frac{1}{\mu_r} \int_{\Sigma_r} d^3 {\bf x} \, \left[ \bfA \cdot
      (\bfn_r \times (\nabla \times \cbA_r)) + (\nabla \times \bfA) \cdot (\bfn_r
      \times \cbA_r )  \right] \, . \nonumber
\end{eqnarray}

Now we have to consider the electric field $\bfE=-\k \bfA$ only on the surfaces
$\Sigma_r$. However, the values of  the electric field $\bfE$
and its curl $\nabla\times\bfE$ are those when the surface is approached from
the inside, denoted by  $\bfE_-$
and $(\nabla\times\bfE)_-$. It is important to realize that in the
above surface integral, $\bfA$
and $\nabla\times\bfA$ multiply vectors that are tangential to the
surface, and hence only the tangential components of $\bfA$
and $\nabla\times\bfA$ contribute to the integral. Hence, we can use
the continuity conditions of the tangential components of $\bfE$ and $\bfH$,
\begin{equation}
  \bfn_r \times \bfE_- =  \bfn_r \times \bfE_+\, , \quad
  \frac{1}{\mu_r}\bfn_r \times (\nabla \times \bfE)_- =
  \frac{1}{\mu_0}\bfn_r \times  (\nabla \times \bfE)_+ \, ,
\end{equation}
to write the source terms as
\begin{eqnarray}
  \label{eq:EM_source_term_2}
  \nonumber
  -\k \int_{V_r} d^3 {\bf x} &&\!\!\!\!\!\!\!\!\!\!\!\!\!\!\!\bfA \cdot \bfP_r =
                                                             \frac{1}{\mu_r} \int_{\Sigma_r} d^3 {\bf x}  \left[ \bfA_-\cdot
      (\bfn_r \times (\nabla \times \cbA_r)) + (\nabla \times \bfA)_-  \cdot (\bfn_r
                                                             \times
                                                             \cbA_r
                                                             )
                                                             \right]\\
                                                         &=&\!\!\!\!
                                                                                              \int_{\Sigma_r} d^3 {\bf x}  \left[  \frac{1}{\mu_r}\bfA_+ \cdot
      (\bfn_r \times (\nabla \times \cbA_r)) +  \frac{1}{\mu_0} (\nabla \times \bfA)_+  \cdot (\bfn_r
      \times \cbA_r )  \right] \, , 
\end{eqnarray}
where the first form applies to $\bfA$ inside the objects and the
second form to $\bfA$ outside the objects. 
There is another advantage of having expressed the latter integrals in terms of the values of $\bfA$ and $\nabla\times\bfA$ when the surfaces are approached from either the outside or the inside of the objects. In the region $V_0$, the field $\bfA\equiv\bfA_0$ is fully determined by its values on the surfaces $\Sigma_r$ and the dielectric function $\epsilon_0$ and permeability $\mu_0$, which are constant across $V_0$. When integrating out $\bfA_0$, in fact, one computes the two-point correlation function of $\bfA_+$ and $(\nabla\times\bfA)_+$ {\it on} the surfaces $\Sigma_r$, and hence the behavior of $\bfA_0$ inside the regions $V_r$ with $r>0$ is irrelevant. Following the same arguments for $\bfA\equiv\bfA_\alpha$ inside the objects, the behavior of $\bfA_r$ outside of region $V_r$ is irrelevant for computing the correlations of $\bfA_-$ and $(\nabla\times\bfA)_-$ {\it on} the surfaces $\Sigma_r$. 
Hence, we can replace in the action $\hat S[\bfA,\{\cbA_r\}]$ the spatially dependent $\epsilon_\bfx$ by $\epsilon_0$ when the coupling of $\bfA_0$ to the surface fields $\cbA_r$ is represented by the second line of Equation~(\ref{eq:EM_source_term_2}), and similarly replace $\epsilon_\bfx$ by $\epsilon_r$ when the coupling of $\bfA_r$ to the surface fields $\cbA_r$ is represented by the first line of Equation~(\ref{eq:EM_source_term_2}).

That this is justified can also be  understood as follows. The field $\bfA_0$ in region $V_0$ can be expanded in a basis of functions that obey the wave equation with $\epsilon_0$. The same can be done for $\bfA_r$ in the interior of each object, i.e., $\bfA_r$ can be expanded in a basis of functions that obey the wave equation with $\epsilon_r$ in $V_r$. For each given set of expansion coefficients in $V_0$ there are corresponding coefficients within each region $V_r$ that are determined by the continuity conditions at the surfaces $\Sigma_r$. The functional integral over $\bfA$ then corresponds to integrating over consistent sets of expansion coefficients that are related by the continuity conditions. The two-point correlations of $\bfA_+$ and  $(\nabla\times\bfA)_+$ on the surfaces $\Sigma_r$ are then fully determined by the integral over the expansion coefficients of $\bfA_0$ in $V_0$ only, and the interior expansion coefficients play no role. Equivalently, the two-point correlations of $\bfA_-$ and  $(\nabla\times\bfA)_-$ on the surfaces $\Sigma_r$ are then fully determined by the integral over the expansion coefficients of $\bfA_r$ in $V_r$ only, and now the exterior expansion coefficients are irrelevant. Hence, in the functional integral, the integration of $\bfA$ can be replaced by $N+1$ integrations over the fields $\bfA_r$, $r=0,\ldots,N$, where each $\bfA_r$ is allowed to extend over unbounded space with the action for a free field in a homogeneous space with $\epsilon_r$, $\mu_r$.
However, it is important that the correct of the two possible forms of the surface integral in Equation~(\ref{eq:EM_source_term_2}) is used.
The  multiple counting of degrees of freedom that results from $N+1$ functional integrations poses no problem since the (formally infinite) factor in the partition function cancels when the Casimir energy is computed from Equation~(\ref{PI_free_energy}).

With this representation, we can write the partition function as a
functional integral over $\bfA$, separately in each region $V_r$,
and the surface fields $\cbA_r$ on body $r$, leading to the partition function 
\be
\label{euclid-z}
\cZ(\kappa) = \prod_{r=0}^N\int\ \cD\bfA_r \prod_{r=1}^N \int \cD\cbA_r
\exp \left[ -\beta \hat S[\{\bfA_r\},\{\cbA_r\}]\right]\, .
\ee
with the action
\begin{eqnarray}
\hat S[\{\bfA_r\},\{\cbA_r\}]&=&-\frac{1}{2}
                                           \sum_{r=0}^N\int_{\mathbb{R}^3}
                                           d^3 {\bf x}\left[  \bfA_r^2 \epsilon_r \k^2 + \frac{1}{\mu_r}
                        (\nabla \times \bfA_r)^2 \right]
  \\ \nonumber
                                       &+&
                                           \sum_{r=1}^N\int_{\Sigma_r} d^3 {\bf x} \, \left[  \frac{1}{\mu_r}\bfA_0
      (\bfn_r \times (\nabla \times \cbA_r)) + \frac{1}{\mu_0} (\nabla \times \bfA_0) (\bfn_r
                                                             \times
                                                             \cbA_r
                                                             )
                                                             \right]\\ \nonumber
&+& \sum_{r=1}^N\int_{\Sigma_r} d^3 {\bf x} \,\left[ \frac{1}{\mu_r}\bfA_r
      (\bfn_r \times (\nabla \times \cbA_r)) + \frac{1}{\mu_r} (\nabla \times \bfA_r) (\bfn_r
                                                             \times
                                                             \cbA_r
                                                             ) \right]
\end{eqnarray}

Now, the fluctuations $\bfA$ can be integrated out easily, noting that
the two point correlation function $\langle \bfA_r(\bfx)
\bfA_{r'}(\bfx')\rangle=0$ for all $r$, ${r'}=0,\ldots,N$ with
$r\neq {r'}$, and for equal-region correlations

\begin{eqnarray}
  \label{eq:E-correlations}
  \langle A_{r,j}(\bfx)  A_{r,k}(\bfx')\rangle &=& {\stackrel{\leftrightarrow}{\cal G}}^{(AA;r)}_{jk}(\bfx,\bfx')
  \\
   \langle (\nabla\times \bfA)_{r,j} (\bfx)  A_{r,k} (\bfx')\rangle &=& \left[ \nabla
                                                              \times {\stackrel{\leftrightarrow}{\cal G}}^{(AA;r)} \right]_{jk}(\bfx, \bfx')
\end{eqnarray}
\vspace{-12pt}
\end{paracol}
\nointerlineskip
\begin{eqnarray}
   \langle  A_{r,j} (\bfx) (\nabla\times \bfA)_{r,k} (\bfx')\rangle &=&  - \left[ 
 {\stackrel{\leftrightarrow}{\cal G}}^{(AA;r)}\times \nabla\right]_{jk}(\bfx,
                                                              \bfx')
                                                              = \left[ \nabla
                                                              \times {\stackrel{\leftrightarrow}{\cal G}}^{(AA;r)}\right]_{jk}(\bfx, \bfx')
\end{eqnarray}
\begin{paracol}{2}
\switchcolumn
\vspace{-12pt}

\begin{eqnarray}
   \langle (\nabla\times \bfA)_{r,j} (\bfx) (\nabla\times \bfA)_{r,k} (\bfx')\rangle &=& - \left[ \nabla\times
         {\stackrel{\leftrightarrow}{\cal G}}^{(AA;r)}\times \nabla\right]_{jk}(\bfx,
                                                              \bfx')
\nonumber \\
&=&  \left[ \nabla\times\nabla
                                                              \times {\stackrel{\leftrightarrow}{\cal G}}^{(AA;r)}\right]_{jk}(\bfx, \bfx')\, 
\end{eqnarray}

\noindent where $\nabla$ always acts on the argument $\bfx$ of ${\stackrel{\leftrightarrow}{\cal G}}^{(AA;r)}$ and
the notation $\nabla \times {\stackrel{\leftrightarrow}{\cal G}}^{(AA;r)}$ means that $\nabla$ acts
column-wise on the tensor ${\stackrel{\leftrightarrow}{\cal G}}^{(AA;r)}$ whereas ${\stackrel{\leftrightarrow}{\cal G}}^{(AA;r)} \times \nabla$
means that $\nabla$ acts row-wise on the tensor ${\stackrel{\leftrightarrow}{\cal G}}^{(AA;r)}$.  We
obtain for the partition function
\begin{eqnarray}
\label{eq:euclid-z-2}
\cZ(\kappa) &=& \prod_{r=1}^N \int \cD\cbA_r
\exp \left[ -\frac{\beta}{2} \left( 
\sum_{r=1}^N\int_{\Sigma_{r}}\!\!\!d^3 {\bf x} \int_{\Sigma_{r}}\!\!\! d^3 {\bf x}' \cbA_r(\bfx) L_{r}(\bfx,\bfx') \cbA_r(\bfx') \right. \right. \nonumber \\
 &+& \left.\left.
\sum_{r,{r'}=1}^N\int_{\Sigma_{r}}\!\!\!d^3 {\bf x} \int_{\Sigma_{{r'}}} \!\!\!d^3 {\bf x}' \cbA_r(\bfx) M_{r{r'}}(\bfx,\bfx') \cbA_{r'}(\bfx')\right)\right]\, .
\end{eqnarray}
with the kernels
\begin{align} 
\label{eq:kernel_M}
L_{r}(\bfx,\bfx') = &\frac{1}{\mu_r^2} \left[ \,
  \nabla \times \nabla \times {\stackrel{\leftrightarrow}{\cal G}}^{(AA;r)}(\bfx,\bfx') (\bfn_r
  \times \cev{\cdot}\,) (\bfn'_r
  \times \vec{\cdot}\,)\right.\nonumber\\
  &+ \left.\nabla \times
  {\stackrel{\leftrightarrow}{\cal G}}^{(AA;r)}(\bfx,\bfx') (\bfn_r 
  \times (\nabla \times \cev{\cdot}\,)) (\bfn'_r
  \times \vec{\cdot}\,) \right. \nonumber \\
  &+ \left. \nabla \times
  {\stackrel{\leftrightarrow}{\cal G}}^{(AA;r)}(\bfx,\bfx') (\bfn_r
  \times \cev{\cdot}\,) (\bfn'_r 
  \times (\nabla' \times \vec{\cdot}\,)) \right. \nonumber\\
&+  \left. {\stackrel{\leftrightarrow}{\cal G}}^{(AA;r)}(\bfx,\bfx')  (\bfn_r 
  \times (\nabla \times \cev{\cdot}\,))  (\bfn'_r 
  \times (\nabla' \times \vec{\cdot}\,)) \right] \nonumber \\
M_{r{r'}}(\bfx,\bfx') = &\frac{1}{\mu_0^2} \,
 \nabla \times \nabla \times {\stackrel{\leftrightarrow}{\cal G}}^{(AA;0)}(\bfx,\bfx') (\bfn_r
  \times \cev{\cdot}\,) (\bfn'_{r'}
  \times \vec{\cdot}\,) \nonumber\\
  &+
\frac{1}{\mu_0\mu_r} \, 
\nabla \times  {\stackrel{\leftrightarrow}{\cal G}}^{(AA;0)}(\bfx,\bfx') (\bfn_r 
  \times (\nabla \times \cev{\cdot}\,)) (\bfn'_{r'}
  \times \vec{\cdot}\,)\nonumber\\
&+ \frac{1}{\mu_0\mu_{r'}} \, \nabla \times  {\stackrel{\leftrightarrow}{\cal G}}^{(AA;0)}(\bfx,\bfx') (\bfn_r
  \times \cev{\cdot}\,) (\bfn'_{r'} 
  \times (\nabla' \times \vec{\cdot}\,))\nonumber\\
&+ \frac{1}{\mu_r\mu_{r'}}  \,  {\stackrel{\leftrightarrow}{\cal G}}^{(AA;0)}(\bfx,\bfx') (\bfn_r 
  \times (\nabla \times \cev{\cdot}\,))  (\bfn'_{r'} 
  \times (\nabla' \times \vec{\cdot}\,))
\end{align}
where ${\stackrel{\leftrightarrow}{\cal G}}^{(AA;r)}(\bfx,\bfx')$ is the free Green function of
Equation~(\ref{eq:G_dyadic}), and the arrow over the placeholder $\cdot$ indicates to which side of the kernel $M$ acts. This notation implies that the derivatives are taken before the kernel is evaluated with $\bfx$ and $\bfx'$ on the surfaces $\Sigma_r$.

\subsection{Hamiltonian Formulation}

The representation of the partition function in the previous subsection  sums over all configurations of the surface fields $\cbA_r$, and the action depends both on  $\cbA_r$ and the tangential part of its curl, which is functionally dependent on $\cbA_r$. Hence, the situation is similar to classical mechanics where the Lagrangian depends on the trajectory $q(t)$ and its velocity $\dot q(t)$. The Lagrangian path integral runs then over all of path $q(t)$ with $\dot q(t)$ determined by the path automatically. To obtain a representation in terms of a space of functions that are defined strictly on the surfaces $\Sigma_r$ only, 
it would be useful to be able to integrate over $\cbA_r$ and its derivatives {\it independently}. In classical mechanics, this is achieved by Lagrange multipliers that lead to a Legendre transformation of the action to its Hamiltonian form. 
Here the situation is similar. To see this, it is important to realize that the bilinear form described by $L_r$ is degenerate on the space of functions over which the functional integral runs, i.e.,
$\int_{\Sigma_{r}}\!\!\!d^3 {\bf x} \int_{\Sigma_{r}}\!\!\! d^3 {\bf x}'
\cbA(\bfx) L_{r}(\bfx,\bfx') \cbA(\bfx')=0$ for all 
$\cbA(\bfx)$ that are regular
solutions of the vector wave equation 
$\nabla \times \nabla \times \cbA + \epsilon_r \mu_r
  \k^2 \cbA  =0$ inside region
$V_r$. With a basis $\{\cbA^{({\rm reg},r)}_\nu(\bfx)\}$ for
this functional space, the elements of $L_r$ can be expressed as
\end{paracol}
\nointerlineskip
\begin{eqnarray}
 \label{eq:L_is_zero}
L_{r}(\nu,\nu') &=&
 \int_{\Sigma_{r}}d^3 {\bf x} \int_{\Sigma_{r}} d^3 {\bf x}'
                         \,\cbA^{({\rm reg},r)}_\nu(\bfx)
                         L_{r}(\bfx,\bfx')
                         \cbA^{({\rm reg},r)}_{\nu'}(\bfx')\nonumber\\
                     &=& \frac{1}{\mu_r^2} \int_{\Sigma_{r}}d^3 {\bf x} \,
                         \left[ \left(\nabla \times  \cbA_{\nu'}(\bfx) \right)
                         \left( \bfn_r \times
                         \cbA^{({\rm reg},r)}_\nu(\bfx) \right)
                         +\cbA_{\nu'}(\bfx)
                         \left( \bfn_r \times \left( \nabla
                         \times  \cbA^{({\rm reg},r)}_\nu(\bfx)\right)\right)
                         \right] \nonumber \\
                     &=& \frac{1}{\mu_r^2}\, \int_{V_r} d^3
                         \bfx \left[
                         \cbA_{\nu'}(\bfx)
                         \left(\nabla\times\nabla\times
                         \cbA^{({\rm reg},r)}_\nu(\bfx)\right) -
                         \cbA^{({\rm reg},r)}_\nu(\bfx) \left(\nabla\times\nabla\times
                          \cbA_{\nu'}(\bfx)\right)
                         \right]
 \nonumber \\
 &=& 0
\end{eqnarray}
\begin{paracol}{2}
\switchcolumn

\noindent where we used the relations of Equation~(\ref{eq:EM_source_term}), and defined
\begin{align}
  \cbA_{\nu'}(\bfx) = & \int_{\Sigma_{r}}d^3 {\bf x}'  \;\left[
\nabla \times {\stackrel{\leftrightarrow}{\cal G}}^{(AA;r)}(\bfx,\bfx') \left( \bfn'_r \times
                         \cbA^{({\rm reg},r)}_{\nu'}(\bfx') \right) \right. \nonumber \\
 &+ \left. {\stackrel{\leftrightarrow}{\cal G}}^{(AA;r)}(\bfx,\bfx') \left( \bfn'_r \times \left( \nabla'
                         \times  \cbA^{({\rm reg},r)}_{\nu'}(\bfx')\right)\right)
\right] \, ,
\end{align}
and made use of the fact that $ \cbA_{\nu'}(\bfx)$ is also a solution of
the vector wave equation inside $V_r$. This implies that the kernel $L_r$ can be
ignored in the above functional integral over regular waves $\cbA_r$ inside the objects.

However, the appearance of the kernel $L_r$ is important in what follows. Let us consider the part of the action $\hat S[\left\{\bfA_r\right\},\left\{\cbA_r\right\}]$ which, after functional integration  over $\bfA_r$, generates the kernel $L_r$. It is given by
\begin{align}
S_r = & -\frac{1}{2} \int_{\mathbb{R}^3} d^3 {\bf x} \;\left[ \bfA_r^2 \epsilon_r \k^2 + \frac{1}{\mu_r}
                        (\nabla \times \bfA_r)^2 \right]
\nonumber \\
& + \frac{1}{\mu_r} \int_{\Sigma_r} d^3 {\bf x} \;\left[ \bfA_r
      (\bfn_r \times (\nabla \times \cbA_r)) + (\nabla \times \bfA_r) (\bfn_r \times \cbA_r ) \right]  \, .
\end{align}

The exponential of this action can be written as a functional integral
over two new vector fields $\bfK_r$ and $\bfK'_r$ that are defined
on the surfaces $\Sigma_r$ and are {\it tangential} to the surfaces,
\end{paracol}
\nointerlineskip
\begin{align}
  \label{eq:EM_rep_kernel_L}
 & \exp(-\beta S_r) \nonumber \\  &=  \cZ_r \oint \cD \bfK_r \cD
                            \bfK'_r
                            \exp \left\{ - \frac{\beta}{2} \frac{1}{\mu_r^2}
                            \int_{\Sigma_r} d^3 {\bf x}
                            \int_{\Sigma_r} d^3 {\bf x}' \left[
                            \bfK_r(\bfx) \cdot \nabla\times\nabla\times
                            {\stackrel{\leftrightarrow}{\cal G}}^{(AA;r)}(\bfx,\bfx') \cdot
                            \bfK_r(\bfx') \right. \right. 
                            \nonumber\\
                            &+ \left.  \left. \bfK_r(\bfx) \cdot
                                \nabla \times {\stackrel{\leftrightarrow}{\cal G}}^{(AA;r)}(\bfx,\bfx') \cdot
                                \bfK'_r(\bfx') +\bfK'_r(\bfx) \cdot
                                \nabla \times {\stackrel{\leftrightarrow}{\cal G}}^{(AA;r)}(\bfx,\bfx')
                               \cdot  \bfK_r(\bfx') 
\right. \right. \nonumber \\                                
                                &+ \left. \left. \bfK'_r(\bfx) \cdot
                                {\stackrel{\leftrightarrow}{\cal G}}^{(AA;r)}(\bfx,\bfx') \cdot
                                \bfK'_r(\bfx')
                            \right] \right.
                            \nonumber\\
  & + \left. \frac{1}{\mu_r} \int_{\Sigma_r} d^3 {\bf x} \,\left[
      \bfA_r \cdot \big(
      (\bfn_r \times (\nabla \times \cbA_r)) - \bfK'_r \big) + (\nabla \times \bfA_r) \cdot \big((\bfn_r
                                                             \times
                                                             \cbA_r
                                                            -\bfK_r) \big) \right] \right\}
\end{align}
\begin{paracol}{2}
\switchcolumn

\noindent where $\cZ_r$ is some normalization coefficient, and we have used $\oint \cD
\bfK_r \cD \bfK'_r$ to indicate that the functional
integral extends only over vector fields that are tangential to the
surface $\Sigma_r$.   This representation shows that
$\bfA_r$ acts as a Lagrange multiplier. Integration over
this field removes the imposed constraints between the {dependent}
tangential fields $\bfn_r\times\cbA_r$, $\bfn_r \times
(\nabla \times \cbA_r)$ by replacing them with the independent
tangential fields $\bfK_r$ and $\bfK'_r$, respectively.

Substituting Equation~(\ref{eq:EM_rep_kernel_L}) for each object into the expression for the partition in Equation~(\ref{euclid-z}), we obtain with $\underline{\bfK}_r=(\bfK_r,\bfK'_r)$
the partition function
\begin{align}
\label{eq:Z_for_2_fields}
\cZ(\kappa) = &\int\ \cD\bfA_0 \prod_{r=1}^N \oint \cD\underline{\bfK}_r 
\\
& \times 
\exp \left[ -\beta S_\text{eff}[\bfA_0,\{\underline{\bfK}_r\}]\right] \exp\left[ -\frac{\beta}{2}\sum_{r=1}^N\int_{\Sigma_r} \!\!\!d^3 {\bf x} \int_{\Sigma_r} \!\!\!d^3 {\bf x}'
\underline{\bfK}_r(\bfx) \hat L_{r}(\bfx,\bfx')\underline{\bfK}_r(\bfx')\right]\nonumber 
\end{align}
with the kernel from Equation~(\ref{eq:EM_rep_kernel_L}),
\begin{equation}
\label{eq:kernel_L_final}
\hat L_{r}(\bfx,\bfx') = \frac{1}{\mu_r^2}
\begin{pmatrix}
\nabla\times\nabla\times {\stackrel{\leftrightarrow}{\cal G}}^{(AA;r)}(\bfx,\bfx') & \nabla\times {\stackrel{\leftrightarrow}{\cal G}}^{(AA;r)}(\bfx,\bfx') \\\nabla\times {\stackrel{\leftrightarrow}{\cal G}}^{(AA;r)}(\bfx,\bfx') &
{\stackrel{\leftrightarrow}{\cal G}}^{(AA;r)}(\bfx,\bfx')
\end{pmatrix}  \, ,
\end{equation}
and with the effective action
\begin{align}
S_\text{eff}[\bfA_0,\{\underline{\bfK}_r\}]=&\frac{1}{2}
\int_{\mathbb{R}^3} d^3 {\bf x}\left[  \bfA_0^2 \epsilon_0 \k^2 + \frac{1}{\mu_0}
                        (\nabla \times \bfA_0)^2\right]
                        \nonumber\\
                      &+\sum_{r=1}^N\int_{\Sigma_r} d^3 {\bf x} \,
                      \left[  \frac{1}{\mu_r}\bfA_0 \bfK'_r+ \frac{1}{\mu_0} (\nabla \times \bfA_0) \bfK_r
                                                             \right]\, ,
\end{align}
where we have integrated out $\bfA_r$ for $r=1,\ldots,N$,
constraining the functional integral over $\cbA_r$ to be replaced
by the substitutions $\bfn_r \times\cbA_r \to \bfK_r$
and $\bfn_r \times (\nabla \times \cbA_r) \to \bfK'_r$. Integrating out $\bfA_0$, finally yields
\begin{align}
\label{eq:euclid-z-3-EM}
\cZ(\kappa) = & \prod_{r=1}^N \oint \cD\underline{\bfK}_r
\exp \left[ -\frac{\beta}{2} \left( 
\sum_{r=1}^N\int_{\Sigma_{r}}\!\!\!d^3 {\bf x} \int_{\Sigma_{r}}\!\!\! d^3 {\bf x}' \, \underline{\bfK}_r(\bfx) \hat L_{r}(\bfx,\bfx') \underline{\bfK}_r(\bfx') \right. \right. \nonumber \\
&+
\left.\left.\sum_{r,{r'}=1}^N\int_{\Sigma_{r}}\!\!\!d^3 {\bf x} \int_{\Sigma_{{r'}}} \!\!\!d^3 {\bf x}' \, \underline{\bfK}_r(\bfx) \hat M_{r{r'}}(\bfx,\bfx') \underline{\bfK}_{r'}(\bfx')\right)\right]\, ,
\end{align}
with the additional kernel
\begin{equation}
\label{eq:kernel_M_2_EM}
\hat M_{r{r'}}(\bfx,\bfx') = 
\begin{pmatrix}
\frac{1}{\mu_0^2} \, \nabla\times\nabla\times {\stackrel{\leftrightarrow}{\cal G}}^{(AA;0)}(\bfx,\bfx') &
\frac{1}{\mu_0 \mu_{r'}} \, \nabla\times {\stackrel{\leftrightarrow}{\cal G}}^{(AA;0)}(\bfx,\bfx') \\
\frac{1}{\mu_0 \mu_r}\, \nabla\times {\stackrel{\leftrightarrow}{\cal G}}^{(AA;0)}(\bfx,\bfx')
&\frac{1}{\mu_r \mu_{r'}} \,  {\stackrel{\leftrightarrow}{\cal G}}^{(AA;0)}(\bfx,\bfx')
\end{pmatrix}  \, .
\end{equation}

It should be noted again that the functional integral in
Equation~(\ref{eq:euclid-z-3-EM}) runs over { tangential} vector fields
$\bfK_r$, $\bfK'_r$ defined on the surfaces $\Sigma_r$ only.
The kernels $\hat L$ and $\hat M$ can be combined into the joint
kernel
\begin{equation}
\hat N_{r{r'}}=\hat L_r \delta_{r{r'}} + \hat
  M_{r{r'}} \, .
\end{equation}

Since $\hat N$ acts in the path integral only on tangential vectors,
the projections of $\hat N$ on the tangent space of the surfaces
$\Sigma_r$ have to be taken. Let $\bft_{r,1}(\bfx)$,
$\bft_{r,2}(\bfx)$ be two tangent vector fields that span the
tangent space of $\Sigma_r$ at $\bfx$. The $4\times 4$ matrix kernels then
become
\end{paracol}
\nointerlineskip
\begin{align}
&\tilde L_{r,mn}(\bfx,\bfx') = \bft_{r,m}(\bfx)\hat L_{r}(\bfx,\bfx') \bft_{r,n}(\bfx')  \nonumber \\
&= \frac{1}{\mu_r}\begin{pmatrix}
   (\bft_{r,m}(\bfx).\nabla)  (\bft_{r,n}(\bfx').\nabla)
   g_r   -  \bft_{r,m}(\bfx) . \bft_{r,n}(\bfx') \nabla
   ^2 g_r  &  - (\bft_{r,m}(\bfx) \times
   \bft_{r,n}(\bfx')) . \nabla g_r\\
    - (\bft_{r,m}(\bfx) \times
   \bft_{r,n}(\bfx')) . \nabla g_r &
   -\frac{1}{\epsilon_r\mu_r\k^2}(\bft_{r,m}(\bfx).\nabla)  (\bft_{r,n}(\bfx').\nabla)
   g_r   +  \bft_{r,m}(\bfx) . \bft_{r,n}(\bfx') g_r 
\end{pmatrix}                                                                                                  
\nonumber
\end{align}
\begin{paracol}{2}
\switchcolumn
\noindent
and
\end{paracol}
\nointerlineskip
\begin{align}
\label{eq:kernel_M_3_EM}
 &\tilde M_{r{r'},mn}(\bfx,\bfx')  = \bft_{r,m}(\bfx)\hat M_{r{r'}}(\bfx,\bfx') \bft_{{r'},n}(\bfx') \nonumber \\
  &= \frac{1}{\mu_0}\begin{pmatrix}
   (\bft_{r,m}(\bfx).\nabla)  (\bft_{r,n}(\bfx').\nabla)
   g_0   -  \bft_{r,m}(\bfx) . \bft_{r,n}(\bfx') \nabla
   ^2 g_0  &  -\frac{\mu_0}{\mu_{r'}} (\bft_{r,m}(\bfx) \times
   \bft_{r,n}(\bfx')) . \nabla g_0\\
    -\frac{\mu_0}{\mu_r} (\bft_{r,m}(\bfx) \times
   \bft_{r,n}(\bfx')) . \nabla g_0 &
   -\frac{\mu_0}{\epsilon_0\mu_r\mu_{r'}\k^2}(\bft_{r,m}(\bfx).\nabla)  (\bft_{r,n}(\bfx').\nabla)
   g_0   +  \frac{\mu_0^2}{\mu_r\mu_{r'}}\bft_{r,m}(\bfx) . \bft_{r,n}(\bfx') g_0 
\end{pmatrix} \nonumber
\end{align}
\begin{paracol}{2}
\switchcolumn
\noindent
which expresses all kernels in terms of tangential and normal
derivatives of the scalar Green function $g_r(|\bfx-\bfx'|)= e^{-\sqrt{\epsilon_r \mu_r}\k |\bfx-\bfx'|}/|\bfx-\bfx'|$. These expressions simplify when an orthonormal
  basis $\bft_{r,1}(\bfx)$,
$\bft_{r,2}(\bfx)$, $\bfn_r(\bfx)=\bft_{r,1}(\bfx) \times \bft_{r,2}(\bfx)$ is used.
The Casimir free energy is then given by
\begin{equation}
\cF = - k_B T \sum_{n=0}^{\infty}\! \!' \log  \det \left[\hat N(\k_n) \hat N_\infty^{-1}(\k_n)\right] \,\label{hamen} ,
\end{equation}
where the determinant runs over all indices, i.e., $\bfx$, $\bfx'$
located {on} the surfaces $\Sigma_r$, and $r$, ${r'}=1,\ldots,N$. The kernel $\hat N_\infty$ is obtained from the kernel $\hat N$ by taking the distance between all bodies to infinity, i.e, by setting $\hat M_{r{r'}} =0$ for all $r\neq {r'}$.  In the following we shall again denote the form of the partition function in Equation~(\ref{eq:euclid-z-2}) as Lagrange representation, and the one of Equation~(\ref{eq:euclid-z-3-EM}) as a Hamiltonian representation.  By a simple computation,  one can verify that the Hamiltonian representation of the Casimir free energy in Equation (\ref{hamen}) is indeed equivalent to the surface formula Equation (\ref{ensurF}).

\section{Application: Derivation of the Lifshitz Theory}

As a simple example to demonstrate the practical application of the surface formulations, we consider two dielectric half-spaces, one covering the region $z\le z_1=0$, with the surface  $\Sigma_1$ and dielectric function $\epsilon_1$ and magnetic permeability $\mu_1$, and the other covering the region $z\ge z_2=H$, with the surface $\Sigma_2$ and dielectric function $\epsilon_2$ magnetic permeability $\mu_2$. We shall consider both the Lagrange and Hamiltonian representation in the following.

\subsection{Lagrange Representation}

We compute the matrix elements of the kernels $L$ and $M$ of Equation~(\ref{eq:kernel_M}) in the basis of transverse vector plane waves, given by
\begin{eqnarray}
\label{eq:basis_planar_vector}
  \bfM_{1,\bfk_\|} &=& \nabla \times \left(e^{-i \bfk_\| \bfx_\|  +
                       p_1 z} \hat \bfz \right) = (-ik_y,ik_x,0) e^{-i \bfk_\| \bfx_\|  +
                       p_1 z} \\
  \bfN_{1,\bfk_\|} &=& \frac{1}{\k}\nabla \times\nabla \times
                       \left(e^{-i \bfk_\| \bfx_\|  + p_1 z} \hat \bfz 
                       \right)
                       = \frac{1}{\k} (-ik_x p_1,-ik_y p_1, k_\|^2) e^{-i \bfk_\| \bfx_\|  + p_1 z}\\
\bfM_{2,\bfk_\|} &=&  \nabla \times \left(e^{-i \bfk_\| \bfx_\|  - p_2
                     (z-H)}\hat \bfz \right) = (-ik_y,ik_x,0) e^{-i
                     \bfk_\| \bfx_\|  - p_2 (z-H)}\\
\bfN_{2,\bfk_\|} &=&   \frac{1}{\k}\nabla \times\nabla \times \left(e^{-i \bfk_\| \bfx_\|  - p_2 (z-H)}\hat \bfz \right)= \frac{1}{\k} (ik_x p_2,ik_y p_2, k_\|^2) e^{-i \bfk_\| \bfx_\|  - p_2 (z-H)}
\end{eqnarray}
with $p_r=\sqrt{\epsilon_r\mu_r\k^2+\bfk_\|^2}$ and the sign of
$z$ is fixed so that the waves are regular inside the
half-spaces. Note that we include here a $z$ dependence to be able to
compute the curl on the surfaces. For the Green tensor, we use the representation
\begin{align}
\label{eq:Green_tensor_plane}
{\stackrel{\leftrightarrow}{\cal G}}^{(AA; r)}({\bf x},{\bf x}') &= \int_\bfq e^{i\bfq (\bfx-\bfx')}
\frac{1/(\epsilon_r\k^2)}{\epsilon_r\mu_r\kappa^2+\bfq^2}
\begin{pmatrix}
  \epsilon_r\mu_r\kappa^2 +q_x^2 & q_x q_y& q_x q_z\\
  q_y q_x & \epsilon_r\mu_r\kappa^2 + q_y^2 & q_y q_z \\
  q_z q_x & q_z q_y & \epsilon_r\mu_r\kappa^2 + q_z^2
\end{pmatrix} \nonumber \\
&\equiv  \int_\bfq e^{i\bfq (\bfx-\bfx')}
\frac{\mu_r  \tilde G_r(\k,\bfq)}{\epsilon_r\mu_r\kappa^2+\bfq^2} \, ,
\end{align}
which yields after the curl operations
\begin{align}
  \label{eq:curl_Green_tensor_plane}
  \nabla \times {\stackrel{\leftrightarrow}{\cal G}}^{(AA; r)}(\bfx,\bfx') &= \int_\bfq e^{i\bfq
                                         (\bfx-\bfx')} \,
                                         \frac{\mu_r}{\epsilon_r\mu_r
                                         \k^2 + \bfq^2}
                                         \begin{pmatrix}
                                           0 & -iq_z & iq_y \\
                                            iq_z & 0 & -iq_x \\
                                            -iq_y &  iq_x& 0 \\
                                          \end{pmatrix} 
                                          \nonumber\\
                                          &\equiv \int_\bfq e^{i\bfq
                                         (\bfx-\bfx')} \,
                                         \frac{\mu_r \tilde G'_r(\bfq)}{\epsilon_r\mu_r
                                         \k^2 + \bfq^2} \\
   \nabla \times \nabla \times {\stackrel{\leftrightarrow}{\cal G}}^{(AA; r)}(\bfx,\bfx') &= \int_\bfq e^{i\bfq
                                         (\bfx-\bfx')} \,
                                         \frac{\mu_r}{\epsilon_r\mu_r
                                         \k^2 + \bfq^2}
                                         \begin{pmatrix}
                                           q_y^2+q_z^2 & -q_xq_y & -q_xq_z \\
                                            -q_x q_y & q_x^2+q_z^2 &
                                            -q_y q_z \\
                                            -q_x q_z &  -q_y q_z& q_x^2+q_y^2 \\
                                           \end{pmatrix} 
                  \nonumber \\                     
                                           &\equiv \int_\bfq e^{i\bfq
                                         (\bfx-\bfx')} \,
                                         \frac{\mu_r \tilde G''_r(\bfq)}{\epsilon_r\mu_r
                                         \k^2 + \bfq^2} \, .
\end{align}

We also need the following expressions for the operators that appear in the
kernels, acting on the basis functions, which are tangential to the
surfaces. On surface $\Sigma_1$ we have
\begin{eqnarray}
\label{eq:op_on_vetor_basis}
\hat z \times \bfM_{1,\bfk_\|} &=& (-ik_x,-ik_y,0) e^{-i \bfk_\|
                                   \bfx_\| } \equiv  \bfu_{m1} e^{-i \bfk_\| \bfx_\| }\\
\hat z \times \bfN_{1,\bfk_\|} &=&  \frac{1}{\k}(ik_y p_1,-ik_x p_1,0) e^{-i \bfk_\| \bfx_\|  } \equiv  \bfu_{n1} e^{-i \bfk_\| \bfx_\| }\\
\hat z \times \nabla \times \bfM_{1,\bfk_\|} &=&   (ik_y p_1,-ik_x p_1,0) e^{-i \bfk_\| \bfx_\| } \equiv  \bfv_{m1} e^{-i \bfk_\| \bfx_\| }\\
\hat z \times \nabla \times \bfN_{1,\bfk_\|} &=&  \frac{1}{\k} (ik_x \epsilon_1\mu_1\k^2, ik_y \epsilon_1\mu_1\k^2,0) e^{-i \bfk_\| \bfx_\| } \equiv  \bfv_{n1} e^{-i \bfk_\| \bfx_\| }
\end{eqnarray}
and similarly on surface $\Sigma_2$, 
\begin{eqnarray}
\label{eq:op_on_vetor_basis}
\hat z \times \bfM_{2,\bfk_\|} &=& (-ik_x,-ik_y,0) e^{-i \bfk_\| \bfx_\|  } \equiv  \bfu_{m2} e^{-i \bfk_\| \bfx_\| }\\
\hat z \times \bfN_{2,\bfk_\|} &=&  \frac{1}{\k}(-ik_y p_2,ik_x p_2,0) e^{-i \bfk_\| \bfx_\|  } \equiv  \bfu_{n2} e^{-i \bfk_\| \bfx_\| }\\
\hat z \times \nabla \times \bfM_{2,\bfk_\|} &=&   (-ik_y p_2,ik_x p_2,0) e^{-i \bfk_\| \bfx_\|  } \equiv  \bfv_{m2} e^{-i \bfk_\| \bfx_\| }\\
\hat z \times \nabla \times \bfN_{2,\bfk_\|} &=&  \frac{1}{\k} (ik_x \epsilon_2\mu_2\k^2, ik_y \epsilon_2\mu_2\k^2,0) e^{-i \bfk_\| \bfx_\|  } \equiv  \bfv_{n2} e^{-i \bfk_\| \bfx_\| }\, .
\end{eqnarray}

It is straightforward to show that the matrix elements of $L_r$ in the above basis all vanish, as the basis functions are regular solutions of the vector wave equation. This observation is in agreement with the above finding that the kernel $L_r$ is degenerate on the space of those solutions.
We proceed with the computation of the elements of kernel $M$.  We find for the  case $r={r'}$,
\begin{align}
\label{eq:EM_kernel_Mdiag_plates}
& M_{rr}(\bfk_\|,\bfk'_\|) = \int_{\Sigma_{r}}d^3 {\bf x}
   \int_{\Sigma_{r}} d^3 {\bf x}' \,
   \begin{pmatrix} \bfM \\ \bfN\end{pmatrix}_{r,\bfk_\|}\!\!\!\!(\bfx)
   M_{rr}(\bfx,\bfx')
    \begin{pmatrix} \bfM \\ \bfN\end{pmatrix}_{r,\bfk'_\|}\!\!\!\!(\bfx')\\
&= \delta(\bfk_\|+\bfk'_\|)\int_{-\infty}^{\infty} \frac{dq_z}{2\pi} 
\left[   \frac{1}{\mu_0^2} \begin{pmatrix} \bfu_{m} \\
    \bfu_{n} \end{pmatrix}_{r,\bfk_\|} \tilde
  G''_0(\bfk_\|,q_z) \begin{pmatrix} \bfu_{m} \\
    \bfu_{n} \end{pmatrix}_{r,-\bfk_\|}  \right.
    \nonumber\\
    &+ \left.
   \frac{1}{\mu_0\mu_r}\begin{pmatrix} \bfv_{m} \\
    \bfv_{n} \end{pmatrix}_{r,\bfk_\|} \tilde
  G'_0(\bfk_\|,q_z) \begin{pmatrix} \bfu_{m} \\
    \bfu_{n} \end{pmatrix}_{r,-\bfk_\|}
  + \frac{1}{\mu_0 \mu_r}\begin{pmatrix} \bfu_{m} \\
    \bfu_{n} \end{pmatrix}_{r,\bfk_\|} \tilde
  G'_0(\bfk_\|,q_z) \begin{pmatrix} \bfv_{m} \\
    \bfv_{n} \end{pmatrix}_{r,-\bfk_\|}  \right. \nonumber 		\\
    &+ \left. \frac{1}{\mu_r^2}\begin{pmatrix} \bfv_{m} \\
    \bfv_{n} \end{pmatrix}_{r,\bfk_\|} \tilde
  G_0(\k,\bfk_\|,q_z) \begin{pmatrix} \bfv_{m} \\
    \bfv_{n} \end{pmatrix}_{r,-\bfk_\|} \right] 
\frac{\mu_0}{\epsilon_0\mu_0
                                         \k^2 +\bfk_\|^2
          +q_z^2}\,
  {e^{iq_z(z-z')}}_{|\,z,z' \to z_r}\nonumber\\
  & =  \delta(\bfk_\|+\bfk'_\|) \frac{\mu_0 \bfk_\|^2}{2p_0}  \begin{pmatrix}
   \frac{\mu_0^2 p_r^2 - \mu_r^2 p_0^2}{(\mu_0\mu_r)^2} & 0\\
    0 & - \frac{\epsilon_0^2 p_r^2-\epsilon_r^2 p_0^2}{\epsilon_0\mu_0}\\
  \end{pmatrix} \nonumber\\
  &\equiv \delta(\bfk_\|+\bfk'_\|)  M_{rr}(\bfk_\|) \nonumber
\end{align}
and for the case $r\neq {r'}$ we get
\clearpage
\end{paracol}
\nointerlineskip
\begin{align}
\label{eq:EM_kernel_Mnondiag_plates}
& M_{r{r'}}(\bfk_\|,\bfk'_\|) = \int_{\Omega_{r}}d^3 {\bf x}
   \int_{\Omega_{{r'}}} d^3 {\bf x}' \,
   \begin{pmatrix} \bfM \\ \bfN\end{pmatrix}_{r,\bfk_\|}\!\!\!\!(\bfx)
   M_{r{r'}}(\bfx,\bfx')
    \begin{pmatrix} \bfM \\ \bfN\end{pmatrix}_{{r'},\bfk'_\|}\!\!\!\!(\bfx')\\
&= \delta(\bfk_\|+\bfk'_\|)\int_{-\infty}^{\infty} \frac{dq_z}{2\pi} 
\left[   \frac{1}{\mu_0^2} \begin{pmatrix} \bfu_{m} \\
    \bfu_{n} \end{pmatrix}_{r,\bfk_\|} \tilde
  G''_0(\bfk_\|,q_z) \begin{pmatrix} \bfu_{m} \\
    \bfu_{n} \end{pmatrix}_{{r'},-\bfk_\|} \right. \nonumber\\
    & + \left.
   \frac{1}{\mu_0\mu_r}\begin{pmatrix} \bfv_{m} \\
    \bfv_{n} \end{pmatrix}_{r,\bfk_\|} \tilde
  G'_0(\bfk_\|,q_z) \begin{pmatrix} \bfu_{m} \\
    \bfu_{n} \end{pmatrix}_{{r'},-\bfk_\|}
  + \frac{1}{\mu_0 \mu_{r'}}\begin{pmatrix} \bfu_{m} \\
    \bfu_{n} \end{pmatrix}_{r,\bfk_\|} \tilde
  G'_0(\bfk_\|,q_z) \begin{pmatrix} \bfv_{m} \\
    \bfv_{n} \end{pmatrix}_{{r'},-\bfk_\|}  \right. \nonumber
    \\
    &+ \left. \frac{1}{\mu_r\mu_{r'}}\begin{pmatrix} \bfv_{m} \\
    \bfv_{n} \end{pmatrix}_{r,\bfk_\|} \tilde
  G_0(\k,\bfk_\|,q_z) \begin{pmatrix} \bfv_{m} \\
    \bfv_{n} \end{pmatrix}_{{r'},-\bfk_\|} \right] 
\frac{\mu_0}{\epsilon_0\mu_0
                                         \k^2 +\bfk_\|^2
          +q_z^2}\,
  {e^{iq_z(-1)^r H}}\nonumber\\
  & =  \delta(\bfk_\|+\bfk'_\|) \frac{\bfk_\|^2}{2\mu_0p_0}  \begin{pmatrix}
   \frac{(\mu_0p_1 - \mu_1 p_0)(\mu_0p_2 - \mu_2 p_0)}{\mu_1\mu_2} & 0\\
    0 & -(\epsilon_0p_1-\epsilon_1 p_0)(\epsilon_0p_2-\epsilon_2 p_0)\frac{\mu_0}{\epsilon_0}\\
  \end{pmatrix} e^{-p_0 H}\, , \nonumber\\
  &\equiv \delta(\bfk_\|+\bfk'_\|)  M_{r{r'}}(\bfk_\|) \, ,\nonumber
\end{align}
\begin{paracol}{2}
\switchcolumn

\noindent where the sign in $e^{i q_z (-1)^r H}$ determines upon
integration over $q_z$ the sign of the terms $\sim i q_z$.
The vanishing of the off-diagonal elements reflects the fact that the two polarizations described by the basis functions $\bfM$ and $\bfN$ do
not couple for planar surfaces. The total kernel $M$ can be written as
\end{paracol}
\nointerlineskip
\begingroup\makeatletter\def\f@size{8}\check@mathfonts
\def\maketag@@@#1{\hbox{\m@th\normalsize\normalfont#1}}%
\begin{align}
  \label{eq:M_EM_plates_final}
  & M(\bfk_\|) =  \nonumber \\
  & \frac{\bfk_\|^2}{2p_0}\begin{pmatrix}
     \frac{\mu_0^2 p_1^2 - \mu_1^2 p_0^2}{\mu_0\mu_1^2} & 0
     &  \frac{(\mu_0p_1 - \mu_1 p_0)(\mu_0p_2 - \mu_2
       p_0)}{\mu_0\mu_1\mu_2} e^{-p_0H} & 0\\
    0 & -\frac{ \epsilon_0^2 p_1^2-\epsilon_1^2
    p_0^2}{\epsilon_0} & 0 & -\frac{(\epsilon_0p_1-\epsilon_1 p_0)(\epsilon_0p_2-\epsilon_2 p_0)}{\epsilon_0}e^{-p_0H}\\
    \frac{(\mu_0p_1 - \mu_1 p_0)(\mu_0p_2 - \mu_2
       p_0)}{\mu_0\mu_1\mu_2} e^{-p_0H}& 0&  \frac{\mu_0^2 p_2^2 - \mu_2^2 p_0^2}{(\mu_0\mu_2^2)} & 0\\
    0 & -\frac{(\epsilon_0p_1-\epsilon_1 p_0)(\epsilon_0p_2-\epsilon_2 p_0)}{\epsilon_0}e^{-p_0H}& 0 & -\frac{ \epsilon_0^2 p_2^2-\epsilon_2^2 p_0^2}{\epsilon_0}
    \end{pmatrix} \nonumber 
\end{align}
\endgroup
\begin{paracol}{2}
\switchcolumn
\noindent

The Casimir free energy is given by
\begin{equation}
\cF = k_B T \sum_{n=0}^{\infty}\! \!' \int
        \frac{d^2\bfk_\|}{(2\pi)^2}\log \det \left[ M
          M^{-1}_\infty  (\bfk_\|) \right]_{\k=\k_n}
      \end{equation}
      in terms of the determinant of the matrix 
      \begin{equation}
        M M_\infty^{-1} (\bfk_\|) = 
        \begin{pmatrix}
 1 & 0 &  \frac{\mu_2}{\mu_1} \frac{\mu_0 p_1-\mu_1 p_0}{\mu_0 p_2+\mu_2 p_0} e^{-p_0 H} & 0 \\
 0 & 1 & 0 & \frac{\epsilon_0 p_1-\epsilon_1 p_0}{\epsilon_0 p_2+\epsilon_2 p_0} e^{-p_0 H}\\
 \frac{\mu_1}{\mu_2} \frac{\mu_0 p_2-\mu_2 p_0}{\mu_0 p_1+\mu_1 p_0}
 e^{-p_0 H}  & 0 & 1 & 0 \\
 0 & \frac{\epsilon_0 p_2-\epsilon_2 p_0}{\epsilon_0 p_1+\epsilon_1 p_0} e^{-p_0 H}  & 0 & 1 \nonumber \\
\end{pmatrix}
 \end{equation}
which has four dimensions due to two sets of basis functions
(polarisations) $\bfM$ and $\bfN$ per
surface. This yields the final result
 \begin{eqnarray}
\cF &=&  k_B T  \sum_{n=0}^{\infty}\! \!' \int
\frac{d\bfk_\|}{(2\pi)^2}\log \left[
  \left( 1 - \frac{(\epsilon_0 p_1 - \epsilon_1 p_0)(\epsilon_0 p_2 -
    \epsilon_2 p_0)}{(\epsilon_0 p_1 + \epsilon_1 p_0)(\epsilon_0 p_2
    + \epsilon_2 p_0)} e^{-2p_0 H} \right) \right. \nonumber \\
&\times&\left. \left( 1 - \frac{(\mu_0 p_1 - \mu_1 p_0)(\mu_0 p_2 -
    \mu_2 p_0)}{(\mu_0 p_1 + \mu_1 p_0)(\mu_0 p_2
    + \mu_2 p_0)} e^{-2p_0 H} \right)
\right]_{\k=\k_n} \label{eq:Plate_energy_EM}
\, .
\end{eqnarray}
\noindent

This result is in agreement with the Lifshitz formula \cite{LifshitzPlates}.

\subsection{Hamiltonian Representation}

Now we derive the Lifshitz expression for the free energy of two dielectric half-spaces in the Hamiltonian representation.
Since the kernels $\hat L$ and $\hat M$ of Equations~(\ref{eq:kernel_L_final}) and (\ref{eq:kernel_M_2_EM}) act on vector fields that are
tangential to the surfaces, we need to compute the matrix elements in
a basis of tangential vectors $\bft_{r,1}(\bfx)$ and
$\bft_{r,2}(\bfx)$ that span the tangent space of surface
  $\Sigma_r$ at position $\bfx$. For a planar surface, one can
  simply set $\bft_{r,1}(\bfx) = \hat \bfx_1$ and  $\bft_{r,2}(\bfx) = \hat \bfx_2$.
For a given pair of positions $\bfx$, $\bfx'$ on the surface and fixed surface indices
$r$, ${r'}$ we obtain the following $4\times 4$ dimensional
matrices
\end{paracol}
\nointerlineskip
\begin{myequation}
\begin{array}{cll}
  && \hat L_{r}(\bfx,\bfx') = \frac{1}{\mu_r^2}
  \begin{pmatrix}
    \hat \bfx_{1}  \nonumber \\ \hat \bfx_{2}
    \end{pmatrix}
\begin{pmatrix}
\nabla\times\nabla\times {\stackrel{\leftrightarrow}{\cal G}}^{(AA; r)}(\bfx,\bfx') & \nabla\times {\stackrel{\leftrightarrow}{\cal G}}^{(AA; r)}(\bfx,\bfx') \\\nabla\times {\stackrel{\leftrightarrow}{\cal G}}^{(AA; r)}(\bfx,\bfx') &
{\stackrel{\leftrightarrow}{\cal G}}^{(AA; r)}(\bfx,\bfx')
\end{pmatrix}
\begin{pmatrix}
    \hat \bfx_{1} \\ \hat \bfx_{2}
  \end{pmatrix}\\
  &\equiv & \frac{1}{\mu_r^2}\begin{pmatrix}
\hat \bfx_{1} .\nabla\times\nabla\times {\stackrel{\leftrightarrow}{\cal G}}^{(AA; r)}(\bfx,\bfx'). \hat \bfx_{1} &
\hat \bfx_{1} .\nabla\times\nabla\times {\stackrel{\leftrightarrow}{\cal G}}^{(AA; r)}(\bfx,\bfx'). \hat \bfx_{2}& \hat \bfx_{1}.\nabla\times
{\stackrel{\leftrightarrow}{\cal G}}^{(AA; r)}(\bfx,\bfx'). \hat \bfx_{1} & \hat \bfx_{1}.\nabla\times {\stackrel{\leftrightarrow}{\cal G}}^{(AA; r)}(\bfx,\bfx'). \hat \bfx_{2}\\
\hat \bfx_{2}.\nabla\times\nabla\times {\stackrel{\leftrightarrow}{\cal G}}^{(AA; r)}(\bfx,\bfx'). \hat \bfx_{1}&
\hat \bfx_{2}.\nabla\times\nabla\times {\stackrel{\leftrightarrow}{\cal G}}^{(AA; r)}(\bfx,\bfx'). \hat \bfx_{2} & \hat \bfx_{2}.\nabla\times
{\stackrel{\leftrightarrow}{\cal G}}^{(AA; r)}(\bfx,\bfx'). \hat \bfx_{1} & \hat \bfx_{2}.\nabla\times {\stackrel{\leftrightarrow}{\cal G}}^{(AA; r)}(\bfx,\bfx'). \hat \bfx_{2} \\
\hat \bfx_{1}.\nabla\times {\stackrel{\leftrightarrow}{\cal G}}^{(AA; r)}(\bfx,\bfx'). \hat \bfx_{1} &\hat \bfx_{1}.\nabla\times {\stackrel{\leftrightarrow}{\cal G}}^{(AA; r)}(\bfx,\bfx'). \hat \bfx_{2} &
\hat \bfx_{1} . {\stackrel{\leftrightarrow}{\cal G}}^{(AA; r)}(\bfx,\bfx'). \hat \bfx_{1} & \hat \bfx_{1} . {\stackrel{\leftrightarrow}{\cal G}}^{(AA; r)}(\bfx,\bfx'). \hat \bfx_{2} \\
\hat \bfx_{2}.\nabla\times {\stackrel{\leftrightarrow}{\cal G}}^{(AA; r)}(\bfx,\bfx') . \hat \bfx_{1}&\hat \bfx_{2}.\nabla\times {\stackrel{\leftrightarrow}{\cal G}}^{(AA; r)}(\bfx,\bfx') . \hat \bfx_{2}&
\hat \bfx_{2} .{\stackrel{\leftrightarrow}{\cal G}}^{(AA; r)}(\bfx,\bfx') . \hat \bfx_{1}& \hat \bfx_{2}
.{\stackrel{\leftrightarrow}{\cal G}}^{(AA; r)}(\bfx,\bfx'). \hat \bfx_{2}
\end{pmatrix}
\nonumber\\
&=&  \frac{1}{\mu_r}\int_\bfq
    \frac{ e^{i\bfq_\|(\bfx_\|-\bfx'_\|)}}{p_r^2 +
    q_z^2}
\begin{pmatrix}
 q_y^2+q_z^2 & -q_x q_y& 0 & -i q_z \\
 -q_x q_y &  q_x^2+q_z^2& i q_z & 0\\
 0 & -i q_z & 1+ \frac{q_x^2}{\epsilon_r \mu_r\k^2}&
 \frac{q_x q_y}{\epsilon_r \mu_r\k^2}\\
 i q_z  & 0 & \frac{q_x q_y}{\epsilon_r \mu_r\k^2} & 1+ \frac{q_y^2}{\epsilon_r \mu_r\k^2}
\end{pmatrix}\nonumber\\
 &=&   \frac{1}{\mu_r}\int_{\bfq_\|}
 \frac{e^{i\bfq_\|(\bfx_\|-\bfx'_\|)}}{2p_r}
\begin{pmatrix}
 q_y^2-p_r^2 & -q_x q_y& 0 & \mp p_r\\
  -q_x q_y & q_x^2 - p_r^2 &   \pm p_r & 0\\
  0 & \mp p_r &  1+ \frac{q_x^2}{\epsilon_r \mu_r\k^2} & \frac{q_x q_y}{\epsilon_r \mu_r\k^2}\\
 \pm p_r  & 0 & \frac{q_x q_y}{\epsilon_r \mu_r\k^2} & 1+ \frac{q_y^2}{\epsilon_r \mu_r\k^2}
\end{pmatrix}\, ,
\nonumber
\end{array}
\end{myequation}
\begin{paracol}{2}
\switchcolumn
\noindent
where we set $\bfx=(\bfx_\|,0)$ and $\bfx=(\bfx_\|,H)$ for surfaces 1
and 2, respectively. We determined the sign of the terms
$\sim i q_z$ from the $q_z$-integration by the observation that $z$, $z'$ have to be taken to the
surface with $z-z'$ staying { inside} the object. The upper (lower)
sign of $p_r$ refers to $r=1$ ($r=2$). 

Analogously, for kernel $M$ we get for the case $r={r'}$
\end{paracol}
\nointerlineskip
\begin{eqnarray}
\nonumber
  \label{eq:M_diag_EM_planar}
  && \hat M_{rr}(\bfx,\bfx') =
  \begin{pmatrix}
    \hat \bfx_{1} \\ \hat \bfx_{2}
    \end{pmatrix}
\begin{pmatrix}
 \frac{1}{\mu_0^2}\nabla\times\nabla\times {\stackrel{\leftrightarrow}{\cal G}}^{(AA;0)}(\bfx,\bfx') &
 \frac{1}{\mu_0\mu_r}\nabla\times {\stackrel{\leftrightarrow}{\cal G}}^{(AA;0)}(\bfx,\bfx') \\
 \frac{1}{\mu_0\mu_r} \nabla\times {\stackrel{\leftrightarrow}{\cal G}}^{(AA;0)}(\bfx,\bfx') &
 \frac{1}{\mu_r^2} {\stackrel{\leftrightarrow}{\cal G}}^{(AA;0)}(\bfx,\bfx')
\end{pmatrix}
\begin{pmatrix}
    \hat \bfx_{1} \\ \hat \bfx_{2}
  \end{pmatrix}\\
&=& \int_\bfq
    \frac{ \mu_0 \, e^{i\bfq_\|(\bfx_\|-\bfx'_\|)}}{p_0^2 +
    q_z^2}
\begin{pmatrix}
 \frac{q_y^2+q_z^2}{\mu_0^2} & -\frac{q_x q_y}{\mu_0^2} & 0 & -\frac{i q_z}{\mu_0\mu_r} \\
 -\frac{q_x q_y}{\mu_0^2} &  \frac{q_x^2+q_z^2}{\mu_0^2}& \frac{i q_z}{\mu_0\mu_r} & 0\\
 0 & -\frac{i q_z}{\mu_0\mu_r} & \frac{1}{\mu_r^2} \left(1+ \frac{q_x^2}{\epsilon_0 \mu_0\k^2}\right)&
 \frac{1}{\mu_r^2}\frac{q_x q_y}{\epsilon_0 \mu_0\k^2}\\
 \frac{i q_z}{\mu_0\mu_r}  & 0 & \frac{1}{\mu_r^2}\frac{q_x q_y}{\epsilon_0 \mu_0\k^2} & \frac{1}{\mu_r^2} \left(1+ \frac{q_y^2}{\epsilon_0 \mu_0\k^2}\right)
\end{pmatrix}\nonumber\\
 &=&  \int_{\bfq_\|}
 \frac{\mu_0 \, e^{i\bfq_\|(\bfx_\|-\bfx'_\|)}}{2p_0}
\begin{pmatrix}
 \frac{q_y^2-p_0^2}{\mu_0^2} & -\frac{q_x q_y}{\mu_0^2} & 0 &
 \frac{\pm p_0}{\mu_0\mu_r} \\
 -\frac{q_x q_y}{\mu_0^2} &  \frac{q_x^2-p_0^2}{\mu_0^2}& \frac{\mp p_0}{\mu_0\mu_r} & 0\\
 0 & \frac{\pm p_0}{\mu_0\mu_r} & \frac{1}{\mu_r^2} \left(1+ \frac{q_x^2}{\epsilon_0 \mu_0\k^2}\right)&
 \frac{1}{\mu_r^2}\frac{q_x q_y}{\epsilon_0 \mu_0\k^2}\\
 \frac{\mp p_0}{\mu_0\mu_r}  & 0 & \frac{1}{\mu_r^2}\frac{q_x q_y}{\epsilon_0 \mu_0\k^2} & \frac{1}{\mu_r^2} \left(1+ \frac{q_y^2}{\epsilon_0 \mu_0\k^2}\right)
\end{pmatrix}
\end{eqnarray}
\begin{paracol}{2}
\switchcolumn

\noindent and for $r\neq{r'}$,
\clearpage
\end{paracol}
\nointerlineskip
\begin{eqnarray}
\nonumber
  \label{eq:M_nondiag_EM_planar}
  && \hat M_{r{r'}}(\bfx,\bfx') =
  \begin{pmatrix}
    \hat \bfx_{1} \\ \hat \bfx_{2}
    \end{pmatrix}
\begin{pmatrix}
 \frac{1}{\mu_0^2}\nabla\times\nabla\times {\stackrel{\leftrightarrow}{\cal G}}^{(AA;0)}(\bfx,\bfx') &
 \frac{1}{\mu_0\mu_{r'}}\nabla\times {\stackrel{\leftrightarrow}{\cal G}}^{(AA;0)}(\bfx,\bfx') \\
 \frac{1}{\mu_0\mu_r} \nabla\times {\stackrel{\leftrightarrow}{\cal G}}^{(AA;0)}(\bfx,\bfx') &
 \frac{1}{\mu_r\mu_{r'}} {\stackrel{\leftrightarrow}{\cal G}}^{(AA;0)}(\bfx,\bfx')
\end{pmatrix}
\begin{pmatrix}
    \hat \bfx_{1} \\ \hat \bfx_{2}
  \end{pmatrix}\\
&=& \int_\bfq
    \frac{ \mu_0 \, e^{i\bfq_\|(\bfx_\|-\bfx'_\|)\mp iq_zH}}{p_0^2 +
    q_z^2}
\begin{pmatrix}
 \frac{q_y^2+q_z^2}{\mu_0^2} & -\frac{q_x q_y}{\mu_0^2} & 0 & -\frac{i q_z}{\mu_0\mu_{r'}} \\
 -\frac{q_x q_y}{\mu_0^2} &  \frac{q_x^2+q_z^2}{\mu_0^2}& \frac{i q_z}{\mu_0\mu_{r'}} & 0\\
 0 & -\frac{i q_z}{\mu_0\mu_r} & \frac{1}{\mu_r\mu_{r'}} \left(1+ \frac{q_x^2}{\epsilon_0 \mu_0\k^2}\right)&
 \frac{1}{\mu_r\mu_{r'}}\frac{q_x q_y}{\epsilon_0 \mu_0\k^2}\\
 \frac{i q_z}{\mu_0\mu_r}  & 0 & \frac{1}{\mu_r\mu_{r'}}\frac{q_x q_y}{\epsilon_0 \mu_0\k^2} & \frac{1}{\mu_r\mu_{r'}} \left(1+ \frac{q_y^2}{\epsilon_0 \mu_0\k^2}\right)
\end{pmatrix}\nonumber\\
 &=&  \int_{\bfq_\|}
 \frac{\mu_0 \, e^{i\bfq_\|(\bfx_\|-\bfx'_\|)}}{2p_0}
\begin{pmatrix}
 \frac{q_y^2-p_0^2}{\mu_0^2} & -\frac{q_x q_y}{\mu_0^2} & 0 &
 \frac{\mp p_0}{\mu_0\mu_{r'}} \\
 -\frac{q_x q_y}{\mu_0^2} &  \frac{q_x^2-p_0^2}{\mu_0^2}& \frac{\pm p_0}{\mu_0\mu_{r'}} & 0\\
 0 & \frac{\mp p_0}{\mu_0\mu_r} & \frac{1}{\mu_r\mu_{r'}} \left(1+ \frac{q_x^2}{\epsilon_0 \mu_0\k^2}\right)&
 \frac{1}{\mu_r\mu_{r'}}\frac{q_x q_y}{\epsilon_0 \mu_0\k^2}\\
 \frac{\pm p_0}{\mu_0\mu_r}  & 0 & \frac{1}{\mu_r\mu_{r'}}\frac{q_x q_y}{\epsilon_0 \mu_0\k^2} & \frac{1}{\mu_r\mu_{r'}} \left(1+ \frac{q_y^2}{\epsilon_0 \mu_0\k^2}\right)
\end{pmatrix} e^{-p_0 H} 
\end{eqnarray}
\begin{paracol}{2}
\switchcolumn

\noindent where again the upper (lower) sign everywhere refers to $r=1$ ($r=2$). 
For the kernel $\hat M$ we determined the sign of the terms
$\sim i q_z$ from the $q_z$-integration by the observation that $z$, $z'$ have to be taken to the
surface with $z-z'$ staying {outside} the object.
When combining the kernels $\hat L$ and $\hat M$ into the joint kernel
$N_{r{r'}}=\hat L_r \delta_{r{r'}} + \hat
M_{rr'}$, it is diagonal in $\bfq_\|$-space with the blocks $N(\bfq_\|)$ on the diagonal given by the $8 \times 8$ matrix shown in Figure~\ref{fig:matrixN}.
\newpage
\begin{figure}[H]	
\includegraphics[scale=0.7]{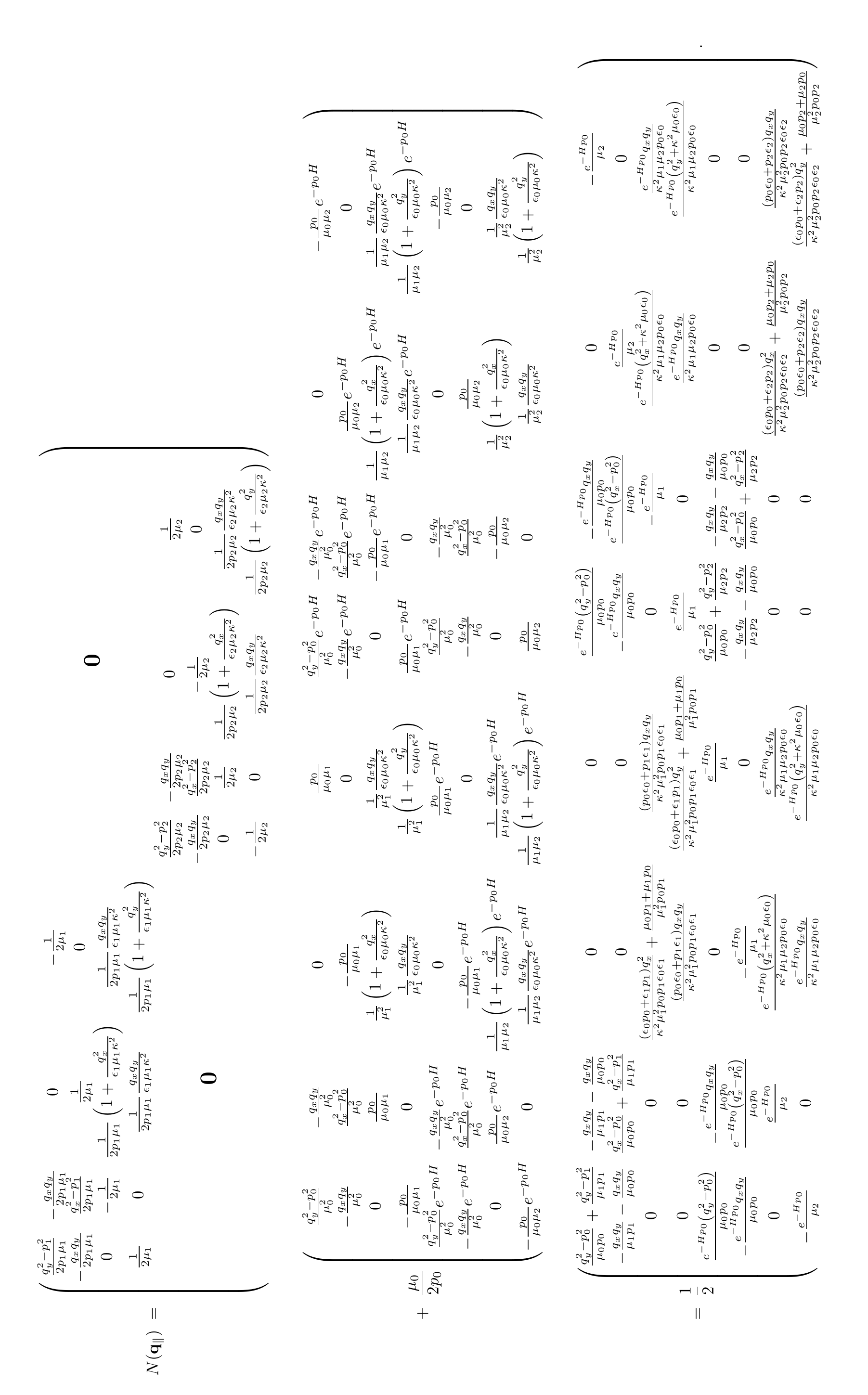}
\caption{Matrix $N(\bfq_\|)$ forming the diagonal blocks of the matrix $N$.\label{fig:matrixN}}
\end{figure}  

The Casimir free energy is given by
\begin{equation}
\cF = k_B T  \sum_{n=0}^{\infty}\! \!'  \int
        \frac{d^2\bfq_\|}{(2\pi)^2}\log \det \left[ N
          N^{-1}_\infty  (\bfq_\|) \right]_{\k=\k_n}
\end{equation}
in terms of the determinant of the above matrix where $N_\infty$ is the matrix with $H\to\infty$, i.e., the matrix $N$ with the
off-diagonal $4\times 4$ blocks $\sim e^{-p_0 H}$ vanishing. This yields the final~result
\begin{eqnarray}
\cF &=&  k_B T  \sum_{n=0}^{\infty}\! \!' \int
\frac{d^2\bfq_\|}{(2\pi)^2}\log \left[
  \left( 1 - \frac{(\epsilon_0 p_1 - \epsilon_1 p_0)(\epsilon_0 p_2 -
    \epsilon_2 p_0)}{(\epsilon_0 p_1 + \epsilon_1 p_0)(\epsilon_0 p_2
    + \epsilon_2 p_0)} e^{-2p_0 H} \right)  \right.\nonumber \\
&\times&\left. \left( 1 - \frac{(\mu_0 p_1 - \mu_1 p_0)(\mu_0 p_2 -
    \mu_2 p_0)}{(\mu_0 p_1 + \mu_1 p_0)(\mu_0 p_2
    + \mu_2 p_0)} e^{-2p_0 H} \right)
\right] \label{eq:Plate_energy_EM}
\, ,
\end{eqnarray}
\noindent
which is again identical to the Lifshitz formula.
Note that in the Hamiltonian approach there is no need to express the kernels in a  basis for the space of functions that are regular solutions of the wave equation inside the objects. However, the number of fields per surface is now doubled compared to the Lagrangian approach, resembling the situation in quantum mechanics where the Hamiltonian path integrals run over the canonical coordinates $q$ and $p$ independently.

\edit{\section{Conclusions}}

\edit{
To date, analytical and purely numerical approaches to compute Casimir interactions have been developed independently, and it remained an open question if and how they are related. Analytical methods build on ideas from scattering theory and hence require an expansion of the Green function and bulk or surface operators in terms of special functions that are solutions of the wave equation. Hence, the very existence of such functions and the convergence of the expansion limit these approaches to sufficiently symmetric problems. Purely numerical approaches, such as that developed in \cite{johnson}, can be applied to basically arbitrary geometries but the numerical effort can be extremely high. Hence, it appeared useful to us to study the relation between these approaches in order to develop methods that can serve as semi-numerical approaches that combine the versatility of the purely numerical approaches with the smaller numerical effort of analytical methods. Hence, we  have presented in this work a new compact derivation of formulas for the Casimir force,  which is based both on bulk and surface operators that also enable  analytical evaluations. This we have demonstrated for the simplest case of two dielectric slabs. Further semi-analytical implementations of our approaches are underway. 

Our Hamiltonian path integral representation is equivalent to the one derived by Johnson et al. as a purely numerical approach using Lagrange multipliers to enforce the boundary conditions in the path integral. Interestingly, the here-presented derivation of this representation from a Lagrangian path integral  demonstrates the relation of this approach to the scattering approach when the T-matrix is defined, as originally by Waterman,  by surface integrals of  regular solutions of the wave equation over the bodies' surfaces \cite{Waterman}. This shows the close connection of these approaches, motivating further research in the direction of new semi-analytical methods to compute Casimir forces.}

\vspace{12pt}

\appendixtitles{yes} 
\appendixstart
\appendix 
\section{Dyadic Green's Functions and Notations} \label{app1}

It is well known \cite{landau} that the expectation values of quantum fields in systems in thermal equilibrium can be expressed, via the Matsubara formalism, in terms of  the analytic continuation to the positive imaginary axis of the appropriate Green's functions. Along the imaginary axis, response functions have a simpler behavior than along the real frequency axis.  For example, it can be shown that the electric and the magnetic permittivities of a magneto-dielectric medium  are real-valued and  positive definite for imaginary frequencies~\cite{landau2}.  In the same manner, retarded Green's functions are real-valued  for positive imaginary frequencies. In this Appendix, we briefly review their definitions and main properties.

To be specific, consider a system consisting of a collection of   possibly spatially dispersive magneto-dielectric bodies in vacuum, occupying the (non-overlapping) regions of space $V_1$, \dots, $V_N$.  We let $V_0$ be the vacuum region that separates the bodies. The   
electromagnetic response of the system is described by the following real-valued and positive definite frequency-dependent electric and magnetic permittivities 
\begin{eqnarray}
\epsilon_{ij}({\bf x},{\bf x}';{\rm i}\,\xi)&=&\sum_{r=0}^N \psi_{r}({\bf x}) \epsilon^{(r)}_{ij}({\bf x},{\bf x}';{\rm i}\,\xi) \psi_{r}({\bf x}')\;, \nonumber\\
\mu_{ij}({\bf x},{\bf x}';{\rm i}\,\xi)&=&\sum_{r=0}^N \psi_{r}({\bf x}) \mu^{(r)}_{ij}({\bf x},{\bf x}';{\rm i}\,\xi) \psi_{r}({\bf x}')\;,\label{epsmu}
\end{eqnarray}
where $\epsilon^{(0)}_{ij}({\bf x},{\bf x}';{\rm i}\,\xi) =\mu^{(0)}_{ij}({\bf x},{\bf x}';{\rm i}\,\xi)=\delta_{ij} \delta^{(3)}({\bf x}-{\bf x}')$ is the permittivity of the vacuum.  For local media,  $\epsilon_{ij}({\bf x},{\bf x}';{\rm i}\,\xi)= \epsilon_{ij}({\bf x}) \delta^{(3)}({\bf x}-{\bf x}')$, and $\mu_{ij}({\bf x},{\bf x}';{\rm i}\,\xi)= \mu_{ij}({\bf x}) \delta^{(3)}({\bf x}-{\bf x}')$.   The principle of microscopic reversibility \cite{eckardt} requires that the permittivities satisfy the symmetry relations:
\begin{eqnarray}
\epsilon_{ij}({\bf x},{\bf x}';{\rm i}\,\xi)&=& \epsilon_{ji}({\bf x}',{\bf x};{\rm i}\,\xi)\;,\nonumber\\
\mu_{ij}({\bf x},{\bf x}';{\rm i}\,\xi)&=& \mu_{ji}({\bf x}',{\bf x};{\rm i}\,\xi)\;.
\end{eqnarray}

The imaginary-frequency Maxwell's Equations are:
\begin{eqnarray}
-{\bf \nabla} \times {\bf E} &=& \kappa ({\bf B}+ 4 \pi \, {\bf M}_{\rm ext})\;,\nonumber\\
{\bf \nabla} \times {\bf H}&=& \kappa ({\bf D}+ 4 \pi \,{\bf P}_{\rm ext})\;,\label{Maxwell}
\end{eqnarray}
where $\kappa=\xi/c$,  $ {\bf P}_{\rm ext}$  and  $ {\bf M}_{\rm ext}$ are the external polarization and magnetization, and 
\begin{eqnarray}
D_i({\bf x})&=&\int_{R^3} d^3 {\bf x}'\,\epsilon_{ij}({\bf x},{\bf x}';{\rm i}\,\xi) \,E_j({\bf x}')\;,\nonumber\\
B_i({\bf x})&=&\int_{R^3} d^3 {\bf x}'\,\mu_{ij}({\bf x},{\bf x}';{\rm i}\,\xi) \,H_j({\bf x}')\;.\label{constit}
\end{eqnarray}

At the boundary separating media $r$ and $s$, the tangential components of ${\bf E}$ and ${\bf H}$ are continuous:
\be
{\hat {\bf n}}\times[{\bf E}^{(r)}-{\bf E}^{(s)}]={\hat {\bf n}}\times[{\bf H}^{(r)}-{\bf H}^{(s)}]=0\;,\label{bc}
\ee
where ${\hat {\bf n}}$ is the unit normal to the boundary.
The dyadic Green's functions ${\cal G}^{(\alpha \beta)}_{ij}({\bf x}-{\bf x}';{\rm i}\,\xi_n)$ are defined such that:
\begin{eqnarray}
{\bf E}({\bf x})&=&\int_{R^3} d^3 {\bf x}'\,\left[ {\stackrel{\leftrightarrow}{\cal G}}^{(EE)}({\bf x},{\bf x}'; {\rm i}\, \xi)\cdot{\bf P}_{\rm ext}({\bf x}') +  {\stackrel{\leftrightarrow}{\cal G}}^{(EH)}({\bf x},{\bf x}'; {\rm i}\, \xi)\cdot{\bf M}_{\rm ext}({\bf x}')\right]\;,\nonumber\\
{\bf H}({\bf x})&=&\int_{R^3} d^3 {\bf x}'\,\left[ {\stackrel{\leftrightarrow}{\cal G}}^{(HE)}({\bf x},{\bf x}'; {\rm i}\, \xi)\cdot{\bf P}_{\rm ext}({\bf x}') +  {\stackrel{\leftrightarrow}{\cal G}}^{(HH)}({\bf x},{\bf x}'; {\rm i}\, \xi)\cdot{\bf M}_{\rm ext}({\bf x}')\right]\;.\label{defgreen0}
\end{eqnarray}

It can be shown \cite{Harrington} that the Green's function satisfies the reciprocity relations:
\be
{ {\cal G}}_{ij}^{(\alpha \beta)}({\bf x},{\bf x}'; {\rm i}\, \xi)=(-1)^{s(\alpha)+s(\beta)}  \, { {\cal G}}_{ji}^{(\beta \alpha)}({\bf x}',{\bf x};{\rm i}\, \xi)\;,\label{recip}
\ee
where  $s(E)=0$ and $s(H)=1$. 

\noindent

In a { local} medium:
\begin{eqnarray}
\epsilon_{ij}({\bf x},{\bf x}';{\rm i}\,\xi)&=& \delta^{(3)}({\bf x}-{\bf x}')\;\epsilon_{ij}({\bf x};{\rm i}\,\xi)\,,\nonumber\\
\mu_{ij}({\bf x},{\bf x}';{\rm i}\,\xi)&=& \delta^{(3)}({\bf x}-{\bf x}')\;\mu_{ij}({\bf x};{\rm i}\,\xi)\,.\label{epsilonmu}
\end{eqnarray}

\textls[-25]{In a local medium, it is possible to express the three Green's function ${\stackrel{\leftrightarrow}{\cal G}}^{(EH)}({\bf x},{\bf x}'; {\rm i}\, \xi)$, ${\stackrel{\leftrightarrow}{\cal G}}^{(HE)}({\bf x},{\bf x}'; {\rm i}\, \xi)$ and ${\stackrel{\leftrightarrow}{\cal G}}^{(HH)}({\bf x},{\bf x}'; {\rm i}\, \xi)$ in terms of the electric Green's function ${\stackrel{\leftrightarrow}{\cal G}}^{(EE)}({\bf x},{\bf x}'; {\rm i}\, \xi)$,~as:}
\begin{eqnarray}
{\stackrel{\leftrightarrow}{\cal G}}^{(HE)}({\bf x},{\bf x}'; {\rm i}\, \xi)&=&
-\frac{1}{\kappa}\,{\stackrel{\leftrightarrow}{\mu}^{-1}\!\!({\bf x},{\rm i}\,\xi)}  \,  {\stackrel{\rightarrow}{\nabla }} \times  {\stackrel{\leftrightarrow}{\cal G}}^{(EE)}({\bf x},{\bf x}'; {\rm i}\, \xi)\;, \label{GHEGEE} \\
{\stackrel{\leftrightarrow}{\cal G}}^{(EH)}({\bf x},{\bf x}'; {\rm i}\, \xi)&=&
-\frac{1}{\kappa}\,  {\stackrel{\leftrightarrow}{\cal G}}^{(EE)}({\bf x},{\bf x}'; {\rm i}\, \xi) \times {\stackrel{\!\!\leftarrow}{\nabla `}}\,{\stackrel{\leftrightarrow}{\mu}^{-1}\!\!({\bf x}',{\rm i}\,\xi)}\;, \label{GEHGEE} \\
{\stackrel{\leftrightarrow}{\cal G}}^{(HH)}({\bf x},{\bf x}'; {\rm i}\, \xi)
&=& \frac{1}{\kappa^2} \, {\stackrel{\leftrightarrow}{\mu}^{-1}\!\!({\bf x},{\rm i}\,\xi)}\;   {\stackrel{\rightarrow}{\nabla }} \times   {\stackrel{\leftrightarrow}{\cal G}}^{(EE)}({\bf x},{\bf x}'; {\rm i} \,\xi)  \times {\stackrel{\!\!\leftarrow}{\nabla `}}\,{\stackrel{\leftrightarrow}{\mu}^{-1}\!\!({\bf x}',{\rm i}\,\xi)} 
\nonumber \\
&-& {4 \pi}\; {\stackrel{\leftrightarrow}{\mu}^{-1}\!\!({\bf x},{\rm i}\,\xi)}\,\delta^{(3)}({\bf x}-{\bf x}')\,, \label{GHHGEE}
\end{eqnarray}
\noindent
where   $  {\stackrel{\!\!\leftarrow}{\nabla `}}$ denotes   derivation w.r.t.  ${\bf x}'$, acting from right.
The electric Green function solves the differential Equation:
\be
{\vec \nabla} \times  \left[ {\stackrel{\leftrightarrow} {\mu}^{-1}\!\!\!({\bf x},{\rm i}\,\xi)} {\vec \nabla} \times  {\stackrel{\leftrightarrow}{\cal G}}^{(EE)}({\bf x},{\bf x}'; {\rm i}\, \xi)\right]+\kappa^2 {\stackrel{\leftrightarrow} {\epsilon}({\bf x},{\rm i}\,\xi)} {\stackrel{\leftrightarrow}{\cal G}}^{(EE)}({\bf x},{\bf x}'; {\rm i}\, \xi) \nonumber \\ 
=-4 \pi \kappa^2\;{\stackrel{\leftrightarrow} 1} \,\delta^{(3)}({\bf x}-{\bf x}')\;.\nonumber
\ee
\noindent

In the case of a homogeneous isotropic medium, this Equation can be solved explicitly:
\be
{\cal G}^{(EE)}_{ij}({\bf x}-{\bf x}'; {\rm i}\, \xi)=-\left(\frac{1}{\epsilon} \frac{\partial^2}{\partial x_i \partial x'_j} + \mu\, \kappa^2\, \delta_{ij} \right)\, g_0( {\bf x}-{\bf x}')\;,\label{greenEEhom}
\ee

From Equations (\ref{GHEGEE})--(\ref{GHHGEE}), one then gets:
\begin{eqnarray}
{\cal G}^{(HH)}_{ij}({\bf x}-{\bf x}' ; {\rm i}\,\xi)&=&-\left(\frac{1}{\mu} \frac{\partial^2}{\partial x_i \partial x'_j} + \epsilon\, \kappa^2\, \delta_{ij} \right)\,  g_0( {\bf x}-{\bf x}')\;,\\
{\cal G}^{(HE)}_{ij}({\bf x}-{\bf x}' ; {\rm i}\, \xi)&=&-\kappa\,\epsilon_{ijk} \frac{\partial}{\partial x_k}\,   g_0( {\bf x}-{\bf x}') \;,\\
{\cal G}^{(EH)}_{ij}({\bf x}-{\bf x}' ; {\rm i}\, \xi)&=& -\kappa\,\epsilon_{ijk} \frac{\partial}{\partial x'_k}\,  g_0( {\bf x}-{\bf x}')\;,\label{greenHHhom}
\end{eqnarray}
\noindent
where
\be
g_0( {\bf x}-{\bf x}')= \frac{e^{-\kappa \sqrt{\epsilon \mu} \vert {\bf x}-{\bf x}' \vert }}{\vert {\bf x}-{\bf x}' \vert}\;.
\ee
is the Green's function of the scalar Helmoltz Equation in free space:
\be
\left(\nabla^2 - \epsilon \,\mu\, \kappa^2 \right)g_0({\bf x}-{\bf x}')= - 4 \pi\,\delta^{(3)}({\bf x}-{\bf x}')\;,
\ee

It is convenient to introduce a compact notation for the fields and the Green's functions. We collect the fields and the sources into  six-rows column-vectors:
\begin{eqnarray}
{ \Phi}&\equiv&\left(\begin{array}{c} { \Phi}^{(E)} ({\bf x})\\  { \Phi}^{(H)}({\bf x})\\ \end{array} \right)=
\left(\begin{array}{c} {\bf E} ({\bf x})\\  {\bf H}({\bf x})\\ \end{array} \right)\;,\;\;\;\;\nonumber\\
{ \cal D}&\equiv&\left(\begin{array}{c} {\bf \cal D}^{(E)} ({\bf x})\\  {\bf \cal D}^{(H)}({\bf x})\\ \end{array} \right)=
\left(\begin{array}{c} {\bf D} ({\bf x})\\  {\bf B}({\bf x})\\ \end{array} \right)\;,\;\;\;\;\nonumber \\
{K}_{\rm ext}&\equiv&\left(\begin{array}{c} {\bf K}_{\rm ext}^{(E)} ({\bf x})\\  {\bf K}_{\rm ext}^{(H)}({\bf x})\\ \end{array} \right)=
\left(\begin{array}{c} {\bf P}_{\rm ext} ({\bf x})\\  {\bf M}_{\rm ext}({\bf x})\\ \end{array} \right)\;.
\end{eqnarray}
\noindent

The defining Equations of the Green's functions  Equation (\ref{defgreen0}) can then be  interpreted  as defining the action of the linear operator $ \hat{\cal G}$ onto the vector ${K}_{\rm ext}$:
\be
{\Phi}= \hat{\cal G} \;{ K}_{\rm ext}\;.
\ee

With this notation, the reciprocity relations Equation (\ref{recip}) are expressed by the operator~Equation:
\be
\hat{\cal G}^{\rm T}=\hat{S}\,\hat{\cal G}\,\hat{S}\;,
\ee
where $\hat{S}$ is the $6 \times 6$ diagonal matrix of elements $S^{(\alpha \beta)}_{ij}=(-1)^{s(\alpha)} \delta_{\alpha \beta} \delta_{ij}$.  The electric and the magnetic permittivities in Equation (\ref{epsmu}) can be both collected into  the permittivity operator $\hat{\epsilon}$. Equation (\ref{epsmu}) becomes:
\be
{\hat \epsilon}=\sum_{r=0}^N \hat{\psi}_r\, \hat{\epsilon}^{(r)} \,\hat{\psi}_r\;,
\ee
where the operator $\hat{\psi}_r$ is 
\be
\hat{\psi}_r \equiv \psi^{(\alpha \beta)}_{r|ij}({\bf x},{\bf x}')=\delta_{\alpha \beta} \,\delta_{ij} \, \delta^{(3)}({\bf x}-{\bf x}') \,\psi_r(\bf{x})\;.
\ee

The constitutive Equations (\ref{constit})  are expressed as:
\be
{\cal D}= \hat{\epsilon}\,\Phi\;.
\ee

For later use, it is convenient to define the polarization operator $\hat{\chi}$:
\be
\hat {\chi}\equiv\frac{1}{4 \pi} \left({\hat \epsilon}- {\hat 1} \right)\;.\label{defchi}
\ee  

Equation (\ref{epsmu}) implies that the polarizability ${\chi}$ of a collection of bodies  is:
\be
\hat{\chi}=\sum_{r=1}^{N} \hat{\chi}^{(r)}\;,
\ee  
where $ \hat{\chi}^{(r)}$ is the polarizability of body $r$. Clearly $ \hat{\chi}^{(r)}$ is supported in the volume $V_r$. This implies:
\be
\hat{\chi}^{(r)} \hat{\chi}^{(s)}=0\;,\;\;\;\;{\rm if\;}r \neq s\;.
\ee

Finally, we define the trace operation ${\rm Tr}$ of any operator ${\hat A}$:
\be
{\rm Tr}\hat{A}= \sum_{i}\sum_{\alpha}\int_{V} d^3 {\bf y} {A}^{(\alpha \alpha)}_{ii}({\bf y})\;.
\ee

\appendix 
\section{Derivation of  Equation (\ref{repGreen3})}  \label{app2}

\noindent

In this Appendix, we derive the volumic and surface representations, Equation (\ref{repGreen1}) and Equation (\ref{repGreen2}), respectively, for    the scattering Green function ${\Gamma}^{(\alpha \beta)}_{ij}({\bf x},{\bf x}' )$.
 
\subsection{Volumic Representation}  \label{app2.1}

Let $\Phi_{\rm ext}= \hat{\cal G}^{(0)} K_{\rm ext}$ be the external field generated  by a certain distribution of external sources $K_{\rm ext}$.  If  one or more   magneto-dielectric bodies are exposed to the field $\Phi_{\rm ext}$, they get polarized, and we let $K_{\rm ind}$ be the induced polarization field.  The polarization field $K_{\rm ind}$ depends linearly on the external  field  $\Phi_{\rm ext}$:
\be
K_{\rm ind}={\hat T}\,\Phi_{\rm ext}\;.\label{defT}
\ee 

The above equation defines  the T-operator of the system. 
The definition of the T-operator implies that the scattered field $\Phi_{\rm sc}$ produced by the bodies is:
\be
\Phi_{\rm sc}=\hat{\cal G}^{(0)} K_{\rm ind}= \hat{\cal G}^{(0)} {\hat T} \Phi_{\rm ext}=  \hat{\cal G}^{(0)} {\hat T}  \hat{\cal G}^{(0)}  K_{\rm ext}\;.
\ee

However, by definition of $\hat{\Gamma}$:
\be
\Phi_{\rm sc}= \hat{\Gamma}\,K_{\rm ext}\;.
\ee

A comparison of the above two equations gives:
\be
\hat{\Gamma}= \hat{\cal G}^{(0)} {\hat T} \; \hat{\cal G}^{(0)}\;,
\ee 
which coincides with Equation  (\ref{repGreen1}).

\noindent

Now we show that the T-operator  can be expressed in terms of the polarizability operator $\hat{\chi}$.
Indeed, by definition of polarizability it holds:
\be
K_{\rm ind}= \hat{\chi} \,\Phi_{\rm tot}\;,
\ee
where the  {\it total} field $\Phi_{\rm tot} =\Phi_{\rm ext}+\Phi_{\rm ind}$ is the sum of the  external field  and of the induced field   $\Phi_{\rm ind}=\hat{\cal G}_0 K_{\rm ind}$  generated by  $K_{\rm ind}$. Therefore:
\be
K_{\rm ind}= \hat{\chi} (\Phi_{\rm ext}+ \Phi_{\rm ind})= \hat{\chi} (\Phi_{\rm ext}+  \hat{\cal G}^{(0)} K_{\rm ind})\;.
\ee

Using Equation (\ref{defT}) to eliminate $K_{\rm ind}$ from the previous equation, we find:
\be
{\hat T}\,\Phi_{\rm ext}= \hat{\chi} (\hat{1}+  \hat{\cal G}^{(0)} \hat{T})\Phi_{\rm ext}\;.
\ee

Since the above equation holds for an arbitrary distribution of sources $K_{\rm ext}$, i.e., for an arbitrary external field $\Phi_{\rm ext}$, it  implies the operator identity:
\be
{\hat T}= \hat{\chi} (\hat{1}+  \hat{\cal G}^{(0)} \hat{T})\;.
\ee

The above relation constitutes an equation of  Lippmann--Schwinger form, which determines the T-operator. Its formal solution  is:
\be
{\hat T} = \frac{1}{1-{\hat \chi}\, \hat{\cal G}^{(0)}}\,{\hat \chi}={\hat \chi}\, \frac{1}{1- \hat{\cal G}^{(0)}\,{\hat \chi}}\;.\label{Top}
\ee

\subsection{Surface Representation}  \label{app2.2}

In this section, we derive the surface formula Equation (\ref{repGreen2}) of the scattering Green's function . The existence of this representation is a direct consequence of the {equivalence  principle} \cite{Harrington} of classical electromagnetism. This principle is applicable to bodies constituted by  homogeneous and isotropic  magneto-dielectric materials whose electric and magnetic permeabilities are of the form\footnote{Bodies that are only piecewise homogeneous can  also be considered  by a slight generalization of the homogeneous case.}:
 \begin{eqnarray}
\epsilon_{ij}({\bf x},{\bf x}';{\rm i}\,\xi)&=& \delta_{ij}\,\delta^{(3)}({\bf x}-{\bf x}')\sum_{r=0}^N \psi_{r}({\bf x}) \epsilon^{(r)}({\rm i}\,\xi) \psi_{r}({\bf x}')\;,\nonumber\\
\mu_{ij}({\bf x},{\bf x}';{\rm i}\,\xi)&=& \delta_{ij}\,\delta^{(3)}({\bf x}-{\bf x}')\sum_{r=0}^N \psi_{r}({\bf x}) \mu^{(r)}({\rm i}\,\xi) \psi_{r}({\bf x}') \;.\label{homog}
\end{eqnarray}

The equivalence principle expresses the scattered field in terms of { fictitious} equivalent currents in a homogeneous medium replacing the scatterer. Let $\Phi$ be the electromagnetic field that solves the Maxwell Equations (\ref{Maxwell}),  with the constitutive  Equation (\ref{constit}) and the permittivities  $\epsilon$ and $\mu$  as in Equation (\ref{homog}), subjected to the  boundary conditions Equation (\ref{bc}) on the surfaces  of N bodies in a vacuum. According to the equivalence principle, there exist {\it tangential surface polarizations} $K_{\rm surf}$ concentrated on the  union $\Sigma$ of the surfaces $\Sigma_r$ of the bodies: 
such that the field $\Phi$ can be expressed as
\begin{eqnarray}
\Phi= \hat{\psi}_0 (\Phi^{(0+)}+  \hat{\cal G}^{(0)} \, K_{\rm surf}) +\sum_{r=1}^{N}\hat{\psi}_r (\Phi^{(r+)} - \hat{\cal G}^{(r)} \, K^{(r)}_{\rm surf}) \;,\label{physurf}
\end{eqnarray}
where $K^{(r)}_{\rm surf}=\hat{\psi}_r K_{\rm surf}$ is the restriction of the surface current to the surface $\Sigma_r$ of the \emph{r}-th body, $\hat{\cal G}^{(r)}$, $r=0,1,\dots N$ is the Green's functions of an infinite homogeneous medium having constant electric and magnetic permittivities equal to $\epsilon^{(r)}({\rm i}\,\xi)$ and $\mu^{(r)}({\rm i}\,\xi)$, respectively, and 
\be
\Phi^{(r+)}=   \hat{\cal G}^{(r)} \,K^{(r)}_{\rm ext}\;,
\ee
is the external field  generated by $K^{(r)}_{\rm ext}$. The Green's functions  $\hat{\cal G}^{(r)}$ are obtained by replacing $\epsilon$
 and $\mu$ by $\epsilon^{(r)}({\rm i}\,\xi)$ and $\mu^{(r)}({\rm i}\,\xi)$, respectively, in Equations (\ref{greenEEhom})--(\ref{greenHHhom}). The equivalence principle shows \cite{Harrington} that the surface currents $K^{(r)}_{\rm surf}$ coincide with the tangential values of the field $\Phi$ on the surface $\Sigma_r$:
 \be
 {K}^{(r)}_{\rm surf}\equiv\left(\begin{array}{c} {\bf K}_{\rm surf}^{(r| E)} ({\bf x})\\  {\bf K}_{\rm surf}^{(r | H)}({\bf x})\\ \end{array} \right)= 
\delta(F_r({\bf x}))\left(\begin{array}{c} \;\;\; \hat{\bf{ n}} \times \hat{\bf H} ({\bf x})\\ - \hat{\bf{ n}} \times \hat{\bf E} ({\bf x})\\ \end{array} \right)\;,\;\;\;\;\;r=1,\dots,N
\ee

In concrete applications of the surface current method to scattering problems, the surface current $ {K}_{\rm surf}$ is unknown.  However, it  can be uniquely determined { a posteriori} by imposing the requirement that  the field $\Phi$ in  Equation (\ref{physurf}) has continuous tangential components across the boundaries of the $N$ bodies. This requirement leads to the following set of Equations:
\be
\hat{\Pi}_r (\Phi^{(0+)}+   \hat{\cal G}^{(0)} \, K_{\rm surf}) =  \hat{\Pi}_r(\Phi^{(r+)} - \hat{\cal G}^{(r)} \, K^{(r)}_{\rm surf})\;,\label{surfeq1}
\ee
where $\hat{\Pi}_r $ is the surface operator, which acts on the field $\Phi({\bf x})$ by projecting it    onto the tangential plane to  $\Sigma_r$  at ${\bf x}$:
\be
{\stackrel{\leftrightarrow}\Pi}_r({\bf x},{\bf x}')=\delta^{(3)}({\bf x}-{\bf x}')\,\delta(F_r({\bf x}))\,\left[ {\stackrel{\leftrightarrow}1}-\hat{\bf{n}}({\bf x}) \otimes  \hat{{\bf n}}({\bf x})\right]\;,\;\;\;\;\;r=1,\dots,N
\ee 

Equation (\ref{surfeq1}) can be recast in the form:
\be
{\hat M} \, K_{\rm surf} = \sum_{r=1}^N \hat{\Pi}_r(  \Phi^{(r+)} -\Phi^{(0+)}) \;,\label{surfeq2}
\ee
where ${\hat M} $ is the {surface} operator
\be
{\hat M} =    \hat{\Pi} \, \hat{\cal G}^{(0)} \,  \hat{\Pi}+\sum_{r=1}^N\hat{\Pi}_r \,   \hat{\cal G}^{(r)}\,\hat{\Pi}_r  \;,\label{defM}
\ee
where $\hat{\Pi}=\sum_{r=1}^N \hat{\Pi}_r\;.$
The operator ${\hat M} $   is invertible \cite{Harrington2}.    Equation (\ref{surfeq2}) then  determines the unique surface current $K_{\rm surf}$ that solves the boundary-value problem:
\be
 K_{\rm surf}=\hat{M}^{-1}\;\sum_{r=1}^N \hat{\Pi}_r(\Phi^{(r+)}- \Phi^{(0+)} ) \label{surf100}
\ee

Note that according to its definition,  $\hat{M}^{-1}$ acts on tangential fields defined on the union $\Sigma$ of the surfaces $\Sigma_r$ of the bodies and returns as a result a tangential polarization field $K_{\rm surf}$  defined on $\Sigma$.
Suppose now that the external sources $K_{\rm ext}$ are  localized in the vacuum region $V_0$:
\be
K_{\rm ext}=  K^{(0)}_{\rm ext}\;.
\ee

Since $\Phi^{(r+)}=0$, Equation (\ref{surf100})  now gives:
\be
 K_{\rm surf}=-\hat{M}^{-1}\; \Phi^{(0+)}=-\hat{M}^{-1}\;   \hat{\cal G}^{(0)}\,K_{\rm ext}^{(0)}\;.\label{surf10}
\ee

According to Equation (\ref{physurf}), the scattered field $ \Phi^{(0)}_{(\rm sc)}=\hat{\psi}_0\,  \Phi_{(\rm sc)}$ in the vacuum region $V_0$ is:
\be
\Phi^{(0)}_{(\rm sc)}=\hat{\cal G}^{(0)}\;K_{\rm surf}=-\hat{\cal G}^{(0)}\;\hat{M}^{-1}\; \hat{\cal G}^{(0)}\,K_{\rm ext}^{(0)}\;.
\ee

The above equation shows that  when $\bf{x}$ and ${\bf x}'$ belong to $V_0$, the scattering Green's function is:
\be
\hat{\Gamma}=-\hat{\cal G}^{(0)}\,\hat{M}^{-1}\,\hat{\cal G}^{(0)}\;.
\ee

We thus see that in the vacuum  outside the bodies  $\hat{\Gamma}$ has precisely the form of \mbox{Equation~(\ref{gammarepgen})}, with 
\be
\hat{\cal{K}}=-\hat{M}^{-1}\;.
\ee

\noindent

\appendix
\section{Proof of the Force Formula Equation (\ref{force1})}  \label{app3}

In this Appendix, we demonstrate the force formula Equation (\ref{force1}).   The simple  proof presented below holds for smooth  kernels    $ {\cal K}_{kl}^{(\rho \sigma)}({\bf y},{\bf y}')$.  Unfortunately,  the kernels $ {\cal K}_{kl}^{(\rho \sigma)}({\bf y},{\bf y}')$ involved in both  the volume and the surface representations of the scattering Green's functions  lack the necessary smoothness.   Indeed the $T$-operator in  Equation (\ref{repGreen1}) is in general discontinuous on the surfaces of the bodies.  As to the surface operator  $\hat{M}$ in Equation (\ref{repGreen2}), it is a singular kernel supported on the surfaces of the bodies.  Fortunately,   it is possible to remedy this difficulty by representing  the  kernel $ {\cal K}_{kl}^{(\rho \sigma)}({\bf y},{\bf y}')$ as the limit of a one-parameter family of   { smooth} real kernels $ {\tilde {\cal K}}_{kl}^{(\rho \sigma)}({\bf y},{\bf y}';\lambda)$ that are supported on an arbitrarily small open neighborhood ${\cal O}$  of   the domain of the original kernel $ {\cal K}_{kl}^{(\rho \sigma)}({\bf y},{\bf y}')$ :
\be
{\cal K}_{kl}^{(\rho \sigma)}({\bf y},{\bf y}')=\lim_{\lambda \rightarrow 0} {\tilde {\cal K}}_{kl}^{(\rho \sigma)}({\bf y},{\bf y}';\lambda)\;.
\ee  

The  approximating kernels $ {\tilde {\cal K}}_{kl}^{(\rho \sigma)}({\bf y},{\bf y}';\lambda)$ can be so defined as to  satisfy the reciprocity relations Equation (\ref{recK})\footnote{For example,  one can set ${\tilde {\cal K}}_{kl}^{(\rho \sigma)}({\bf y},{\bf y}';\lambda)= \lambda^{-6} \int_{V} d^3 {\bf x} \int_{V} d^3 {\bf x}' f(({\bf y}-{\bf x})/\lambda)  f(({\bf y}'-{\bf x}')/\lambda) {\cal K}_{kl}^{(\rho \sigma)}({\bf x},{\bf x}')$, where $f({\bf x})$ is any  rotationally invariant,  smooth non-negative function,   supported  in a ball of unit radius centered  in the origin,  such that $\int d^3{\bf x} f({\bf x})=1$.}:
\be
\tilde {{\cal K}}_{kl}^{(\rho \sigma)}({\bf y},{\bf y}';\lambda) = (-1)^{s(\rho)+s(\sigma)} \tilde{{\cal K}}_{lk}^{(\sigma \rho)}({\bf y}',{\bf y};\lambda) \;,\label{recK2}
\ee
where we recall that $s(\alpha)$ is defined such that $s(E)=0$ and $s(H)=1$.
It thus holds:
\be
{\Gamma}^{(\alpha \beta)}_{ij}({\bf x},{\bf x}' )=\lim_{\lambda \rightarrow 0} {\tilde {\Gamma}}^{(\alpha \beta)}_{ij}({\bf x},{\bf x}' ;\lambda)\;,
\ee
where we set
\be
\tilde{{\Gamma}}^{(\alpha \beta)}_{ij}({\bf x},{\bf x}' ;\lambda) \equiv \int_{\cal O}  d^3 {\bf y} \int_{\cal O} d^3 {\bf y}'   {\cal G}^{(\alpha \rho;0)}_{ik}({\bf x}-{\bf y} )   {\tilde {\cal K}}_{kl}^{(\rho \sigma)}({\bf y},{\bf y}';\lambda)   {\cal G}_{lj}^{(\sigma \beta;0)}({\bf y}'-{\bf x}')\;.\label{tildeGreen}
\ee 

For brevity, below we shall omit writing the parameter $\lambda$, and we shall consider the $\lambda \rightarrow 0$ limit only at the end.
In the first step, we use reciprocity relations satisfied by the free-space Green function to rewrite Equation  (\ref{tildeGreen}) as:
\vspace{9pt}

\end{paracol}
\nointerlineskip
\appendix
\be
\tilde{{\Gamma}}^{(\alpha \beta)}_{ij}({\bf x},{\bf x}' )= \int_{\cal O}  d^3 {\bf y} \int_{\cal O} d^3 {\bf y}'  {\cal G}^{(\alpha \rho;0)}_{ik}({\bf x}-{\bf y} )    {\tilde {\cal K}}_{kl}^{(\rho \sigma)}({\bf y},{\bf y}')  (-1)^{s(\sigma)+s(\beta)} {\cal G}_{jl}^{(\beta \sigma;0)}({\bf x}'-{\bf y}')\;.\label{repGreen4}
\ee 
\begin{paracol}{2}
\switchcolumn
Next, one notes that, by virtue of the reciprocity relations Equation (\ref{recK2}),  the real kernel $ (-1)^{s(\sigma)} \tilde{{\cal K}}_{kl}^{(\rho \sigma)}({\bf y},{\bf y}')$ is symmetric, and therefore, it can be diagonalized:

\be
(-1)^{s(\sigma)} \tilde{{\cal K}}_{kl}^{(\rho \sigma)}({\bf y},{\bf y}')= \sum_m w_m K^{(\rho)}_{m|k}({\bf y})\,K^{(\sigma)}_{m | l} (\bf{y}')\;,\label{diagtildeK}
\ee
where the eigenvalues $w_m$ are non-vanishing real numbers, and the eigenvectors $ K^{(\rho)}_{m|k}({\bf y})$ are real smooth fields supported in ${\cal O}$. By plugging the above expansion  into \mbox{Equation (\ref{repGreen4})}, we see that $\tilde{{\Gamma}}^{(\alpha \beta)}_{ij}({\bf x},{\bf x}' )$ can be expressed as:
\be
\tilde{{\Gamma}}^{(\alpha \beta)}_{ij}({\bf x},{\bf x}' )=(-1)^{s(\beta)} \sum_m w_m  \Phi^{(\alpha)}_{m|i}({\bf x})\,\Phi^{(\beta)}_{m | j} (\bf{x}')\;,\label{repGamma2}
\ee
where $ \Phi^{(\alpha)}_{m|i}({\bf x})$ are the real fields:
\be
 \Phi^{(\alpha)}_{m|i}({\bf x}) =\int_{\cal O}  d^3 {\bf y}\,   {\cal G}^{(\alpha \rho;0)}_{ik}({\bf x}-{\bf y} ) K^{(\rho)}_{m|k}({\bf y})\;.\label{fields40}
\ee

Now one notes that   the fields $\Phi^{(\alpha)}_{m|i}({\bf x})$ are  well defined in all space, and by construction they satisfy the following euclidean-time Maxwell Equations:
\begin{eqnarray}
-{\bf \nabla} \times {\bf E}_m &=& \kappa ({\bf H}_m+ 4 \pi \, {\bf M}_m)\;,\\
{\bf \nabla} \times {\bf H}_m &=& \kappa ({\bf E}_m+ 4 \pi \,{\bf P}_m)\;,
\end{eqnarray}
where we set ${ \Phi}_m \equiv ({\bf E}_m, {\bf H}_m)$, ${\bf K}_m \equiv ({\bf P}_m, {\bf M}_m)$ and $\kappa= \xi_n/c$.
\textls[-25]{When the expansion \mbox{Equation (\ref{repGamma2})}  is substituted into Equation (\ref{stressGamma}),  we find for the dyad ${\stackrel{\leftrightarrow}{\Theta}}$ the following~expression:}
\be
{\stackrel{\leftrightarrow}{\Theta}} = 2 k_B T \left.\sum_{n=0}^{\infty}\right.\!'  \sum_m w_m \left[{\stackrel{\leftrightarrow}{\Theta}}_{m}^{(EE)} - {\stackrel{\leftrightarrow}{\Theta}}_{m}^{(HH)}  \right]\;, \label{calTrep}
\ee
where
\begin{eqnarray}
{\stackrel{\leftrightarrow}{\Theta}}_{m}^{(EE)} &=& \frac{1}{4 \pi} \left[{\bf E}_m \otimes {\bf E}_m- \frac{E^2_m}{2} {\stackrel{\leftrightarrow} 1} \right]\;,\\
{\stackrel{\leftrightarrow}{\Theta}}_{m}^{(HH)}&=&  \frac{1}{4 \pi} \left[{\bf H}_m \otimes {\bf H}_m- \frac{H^2_m}{2} {\stackrel{\leftrightarrow} 1}\right]\;.
\end{eqnarray}

It is worth noticing that the minus sign that multiplies ${\stackrel{\leftrightarrow}{\Theta}}_{m}^{(HH)}$ in Equation (\ref{calTrep}) is a direct consequence of the factor $(-1)^{s(\beta)}$ in the r.h.s. of Equation (\ref{repGamma2}). At this point, we use the divergence theorem to convert the surface integral in Equation (\ref{force0}), giving the force ${\bf F}^{(r)}$, into a volume integral. That gives:
\be
{\bf F}^{(r)}=\int_{{\cal O}_r} d^3 {\bf y}\, {\bf \nabla} \cdot  {\stackrel{\leftrightarrow}{\Theta}}=2 k_B T \left.\sum_{n=0}^{\infty}\right.\!'  \sum_m w_m \, \int_{{\cal O}_r} d^3 {\bf y}{\bf \nabla} \cdot  \left[{\stackrel{\leftrightarrow}{\Theta}}_{m}^{(EE)} - {\stackrel{\leftrightarrow}{\Theta}}_{m}^{(HH)}  \right]\;. \label{force30}
\ee 
where ${\cal O}_r$ is the portion of ${\cal O}$ included within the surface $S_r$.
By using standard identities of vector calculus, and the Maxwell Equations satisfied by the fields   $({\bf E}_m, {\bf H}_m)$ one finds:
$$
{\bf \nabla} \cdot  \left[{\stackrel{\leftrightarrow}{\Theta}}_{m}^{(EE)} - {\stackrel{\leftrightarrow}{\Theta}}_{m}^{(HH)}  \right]=\kappa {\bf E}_m \times {\bf M}_m- {\bf E}_m ({\bf \nabla} \cdot {\bf P}_m)+\kappa {\bf H}_m \times {\bf P}_m + {\bf H}_m( {\bf \nabla} \cdot {\bf M}_m)\;
$$
\be
=({\bf \nabla} \times {\bf H}_m) \times {\bf M}_m- {\bf E}_m ({\bf \nabla} \cdot {\bf P}_m)-({\bf \nabla} \times {\bf E}_m) \times {\bf P}_m+ {\bf H}_m ({\bf \nabla} \cdot {\bf M}_m)\;.
\ee

Upon substituting the r.h.s. of the above equation into the r.h.s. of Equation (\ref{force30}) we~get
\be
{\bf F}^{(r)}_i=2 k_B T \left.\sum_{n=0}^{\infty}\right.\!'  \sum_m w_m \, \int_{{\cal O}_r} d^3 {\bf y} \left[ P_{m | j} \partial_i E_{m| j}- M_{m | j} \partial_i H_{m| j}\right]\;.
\ee

If  the fields ${\bf E}_m$ and ${\bf H}_m$ are expressed in terms of the sources ${\bf P}_m$ and ${\bf M}_m$, via \mbox{Equation (\ref{fields40})} one can recast the r.h.s. of the above equation in the form:
\vspace{9pt}

\end{paracol}
\nointerlineskip
$$
{\bf F}^{(r)}_i=2 k_B T \left.\sum_{n=0}^{\infty}\right.\!'  \sum_m w_m \, \int_{{\cal O}_r} d^3 {\bf y} \int_{\cal O}  d^3 {\bf y}'
(-1)^{s(\alpha)}  K^{(\alpha)}_{m | j}({\bf y}) \frac{\partial}{\partial {y_i}} {\cal G}^{(\alpha \beta;0)}_{jk}({\bf y}-{\bf y}') K^{(\beta)}_{m| k}({\bf y}') 
$$
\be
=2 k_B T \left.\sum_{n=0}^{\infty}\right.\!'    \, \int_{{\cal O}_r} d^3 {\bf y} \int_{\cal O}  d^3 {\bf y}'
{\tilde K}^{(\beta \alpha)}_{kj}({\bf y}',{\bf y};{\lambda})  \frac{\partial}{\partial {y_i}} {\cal G}^{(\alpha \beta;0)}_{jk}({\bf y}-{\bf y}')  \;,\label{forcetildeK}
\ee
\begin{paracol}{2}
\switchcolumn
\noindent where in the last passage, we made use of Equation (\ref{diagtildeK}), and restored the dependence of ${\tilde K}$ on the parameter $\lambda$. By taking the limit $\lambda \rightarrow 0$ of the r.h.s.,   we  see that Equation (\ref{forcetildeK}) reproduces the formula for force in Equation (\ref{force1}).

\appendix
\section{Partial Wave Expansion and the Scattering Matrix}  \label{app4}

The Green's function ${\stackrel{\leftrightarrow}{\cal G}}^{(\alpha \beta;0)}({\bf x},{\bf x}')$ admits an expansion in partial waves in any coordinate system in which the vector Helmoltz Equation is separable.  When spherical coordinates are used,   the partial-wave expansion  takes the form of an infinite series over spherical multipoles:
\begin{eqnarray}
{\stackrel{\leftrightarrow} {\cal G}}^{(\alpha \beta;0)}({\bf x},{\bf x}')&=&(-1)^{s(\beta)} \,\lambda\,\sum_{ilm}  \left[ \theta(|{\bf x}|-|{\bf x}'|) { \Phi}^{(\alpha | {\rm out})}_{ilm}({\bf x}) \otimes { \Phi}^{(\beta | {\rm reg})}_{il-m}({\bf x}') \right. \nonumber \\
&+&\left.\theta(|{\bf x}'|-|{\bf x}|) { \Phi}^{(\alpha | {\rm reg})}_{ilm}({\bf x}) \otimes { \Phi}^{(\beta | {\rm out})}_{il-m}({\bf x}')  \right]\;,\label{pwexp}
\end{eqnarray}
\noindent
where $\lambda=-4 \pi \, \kappa^3$,  $i=M,N$ and $(lm)$ are, respectively, polarization and multipole indices. The partial waves ${ \Phi}^{(\alpha |{\rm out/reg})}_{ilm}({\bf x})$ are defined as follows. For the electric field, they are:  \cite{Rahi}:
\begin{eqnarray}
 { \Phi}^{(E | {\rm out/reg})}_{Mlm}({\bf x})&=&\frac{1}{\sqrt{l(l+1)}}\,\nabla \times {\bf x}\,\phi_{lm}^{({\rm out/reg})}({\bf x},  \kappa )  \;,\\
{ \Phi}^{(E | {\rm out/reg})}_{Nlm}({\bf x})&=& \frac{{\rm i}}{\kappa } \nabla \times  { \Phi}^{(E | {\rm out/reg})}_{Mlm}({\bf x},  \kappa)\;,\label{pwrel1}
\end{eqnarray}
where $\phi_{lm}^{({\rm reg/out})}({\bf x},\kappa )$ are the following regular and the outgoing spherical  waves:
\begin{eqnarray}
\phi_{lm}^{({\rm reg})}({\bf x},\kappa )&=& i_l(\kappa |{\bf x}|)Y_{lm}(\hat{\bf x})\;,\\
\phi_{lm}^{({\rm out})}({\bf x},\kappa )&=& k_l(\kappa |{\bf x}|)Y_{lm}(\hat{\bf x})\;.
\end{eqnarray}

Here, $i_l(z)=\sqrt{\pi/2 z} I_{l+1/2}(z)$ is the modified spherical Bessel function of the first kind, and $k_l(z)=\sqrt{\pi/2 z} K_{l+1/2}(z)$ is the modified spherical Bessel function of the third kind. Notice the relation:
\be
{ \Phi}^{(E | {\rm out/reg})}_{Mlm}({\bf x})= \frac{{\rm i}}{\kappa  } \nabla \times  { \Phi}^{(E | {\rm out/reg})}_{Nlm}({\bf x})\;.\label{pwrel2}
\ee

According to Maxwell Equations,  the  magnetic partial waves ${ \Phi}^{(H | {\rm reg/out})}_{ilm}({\bf x})$  are obtained by taking a curl of the electric waves:
 \be
 { \Phi}^{(H | {\rm reg/out})}_{ilm}({\bf x}) \equiv -\frac{1}{\kappa  } \nabla \times  { \Phi}^{(E | {\rm reg/out})}_{ilm}({\bf x}) \;.\label{Hpw}
\ee

However, using  the two relations  Equations (\ref{pwrel1}) and (\ref{pwrel2}), one finds:
 \be
 { \Phi}^{(H | {\rm reg/out})}_{ilm}({\bf x}) =  \,    {\rm i}\,{ \Phi}^{(E | {\rm reg/out})}_{\tau(i)lm}({\bf x})\;.\label{relHE}
\ee
where   $\tau(M)=N$, $\tau(N)=M$. 

\noindent

Next, we define the scattering matrix ${\cal T}^{(r)}$ of an { isolated} object $r$ placed in vacuum. Let    ${\stackrel{\leftrightarrow} {\Gamma}}^{(\alpha \beta)}_r({\bf x},{\bf x}')$ be the scattering part of the  Green's function of body $r$ in isolation. The scattering matrix ${\cal T}^{(r)}$ is defined such that at  all points $({\bf x},{\bf x}')$ lying { outside} a sphere $S^{(r)}$  containing the body in its interior and centered at the point ${\bf X}_r$,  ${\stackrel{\leftrightarrow} {\Gamma}}^{(\alpha \beta)}_r({\bf x},{\bf x}')$ has the partial wave expansion:
\be
{\stackrel{\leftrightarrow}{\Gamma}}^{(\alpha \beta)}_r({\bf x},{\bf x}')=(-1)^{s(\beta)}
\lambda \sum_{plm} \sum_{p' l' m'}   { \Phi}^{(\alpha | {\rm out})}_{plm}({\bf x}-{\bf X}_r) \otimes { \Phi}^{(\beta | {\rm out})}_{p'l'-m'}({\bf x}'-{\bf X}_r) \;{\cal T}^{(r)}_{plm,p'l'm'}\;,\label{tmatdef}
\ee

The  fact that all Green's functions ${\stackrel{\leftrightarrow}{\Gamma}}^{(\alpha \beta)}_r({\bf x},{\bf x}')$ can be expressed in terms of a single scattering matrix ${\cal T}^{(r)}$ follows from Equation (\ref{Hpw}), together with the following identities satisfied  by  ${\stackrel{\leftrightarrow}{\Gamma}}^{(\alpha \beta)}_r({\bf x},{\bf x}')$ at all points in the vacuum outside the body, which in turn result from the   identities of Equations (\ref{GHEGEE})--(\ref{GHHGEE}) satisfied by the full Green's functions ${\stackrel{\leftrightarrow}{G}}^{(\alpha \beta)}({\bf x},{\bf x}')$  in the presence of body $r$:
\begin{eqnarray}
{\stackrel{\leftrightarrow}{\Gamma}}^{(HE)}_r({\bf x},{\bf x}'; {\rm i}\, \xi)&=&
-\frac{1}{\kappa}\,  {\stackrel{\rightarrow}{\nabla }} \times  {\stackrel{\leftrightarrow}{\Gamma}}^{(EE)}_r({\bf x},{\bf x}'; {\rm i}\, \xi)\;, \nonumber \\
{\stackrel{\leftrightarrow}{\Gamma}}^{(EH)}_r({\bf x},{\bf x}'; {\rm i}\, \xi)&=&
- \frac{1}{\kappa}\,   {\stackrel{\leftrightarrow}{\Gamma}}^{(EE)}_r({\bf x},{\bf x}'; {\rm i}\, \xi) \times {\stackrel{\!\!\leftarrow}{\nabla `}}\;, \nonumber\\
{\stackrel{\leftrightarrow}{\Gamma}}^{(HH)}_r({\bf x},{\bf x}'; {\rm i}\, \xi)
&=& \frac{1}{\kappa^2}     {\stackrel{\rightarrow}{\nabla }}\times  {\stackrel{\leftrightarrow}{\Gamma}}^{(EE)}_r({\bf x},{\bf x}'; {\rm i} \,\xi) \times {\stackrel{\!\!\leftarrow}{\nabla `}} \,.  
\end{eqnarray}

Now we prove that the  scattering matrix  is given by the matrix elements of the operator $\hat{{\cal K}}_r$ of body $r$ defined in Equation (\ref{gammarepgenis}), taken between two regular partial waves.  To see this, one notes that for any two points  $({\bf x},{\bf x}')$  {outside} the sphere $S^{(r)}$, it is legitimate to replace $\hat{\cal G}^{(0)}$ in Equation (\ref{gammarepgenis}) by its partial wave expansion Equation (\ref{pwexp}). This substitution results in the expansion:
\begin{eqnarray}
&&\!\!\!\!{\stackrel{\leftrightarrow}{\Gamma}}^{(\alpha \beta)}_r({\bf x},{\bf x}')=- (-1)^{s(\beta)}\lambda^2 \sum_{plm} \sum_{p' l' m'}  { \Phi}^{(\alpha | {\rm out})}_{plm}({\bf x}-{\bf X}_r) \otimes { \Phi}^{(\beta | {\rm out})}_{p' l'-m'}({\bf x}'-{\bf X}_r) \nonumber \\
\times \sum_{\mu \nu} &&\!\!\!\!\!\!\!\! (-1)^{s(\mu)}\int_{V_r} d^3 {\bf y}  \int_{{V}_r} d^3 {\bf y}'  { \Phi}^{(\mu | {\rm reg})}_{pl-m}({\bf y}-{\bf X}_r) \cdot  {\stackrel{\leftrightarrow}{\cal K}}^{(\mu \nu)}_r ({\bf y},{\bf y}') \cdot  { \Phi}^{(\nu | {\rm reg})}_{p' l' m'}({\bf y}'-{\bf X}_r)\;.  \label{scatGsurf}
\end{eqnarray}
\noindent

A comparison of Equation (\ref{tmatdef}) with Equation (\ref{scatGsurf}) gives the desired formula:
\begin{eqnarray}
{\cal T}^{(r)}_{plm,p'l'm'}&=& \sum_{\mu \nu} (-1)^{s(\mu)} \lambda\int_{{V}_r} d^3 {\bf y}  \int_{{V}_r} d^3 {\bf y}' \;  \nonumber \\
&\times&{ \Phi}^{(\mu | {\rm reg})}_{pl-m}({\bf y}-{\bf X}_r) \cdot  {\stackrel{\leftrightarrow}{\cal K}}^{(\mu \nu)}_r ({\bf y},{\bf y}') \cdot  { \Phi}^{(\nu | {\rm reg})}_{p' l' m'}({\bf y}'-{\bf X}_r)\;,\label{tmatK}
\end{eqnarray}
\noindent
which indeed shows that  ${\cal T}^{(r)}_{plm,p'l'm'}$  is the matrix element of the operator $\hat{{\cal K}}_r$ between two partial waves. Note that in the $T$-operator approach, ${\cal T}^{(r)}_{plm,p'l'm'}$  is expressed by an integral of the body's $T$-operator ${\hat T}_r$ over the body's volume $V_r$, while in the surface approach it is an integral of the surface operator $-{\hat M}_r^{-1}$ over the boundary $\Sigma_r$ of the body. 
\noindent

Now we define the translation matrices ${\cal U}^{(ij)}(d)$.  Consider two points ${\bf X}_1$  and  ${\bf X}_2$ in space   that differ by a displacement of magnitude $d$ along the $z$-axis:
\be
{\bf X}_2-{\bf X}_1=d\,{\hat {\bf z}}\;.
\ee 

The translation matrix ${\cal U}^{(21)}(d)$ is defined such that at all points ${\bf x}$ whose distances from ${\bf X}_1$ and   ${\bf X}_1$ are both smaller than $d$  (i.e., $|{\bf x}-{\bf X}_2| <  d$ and $|{\bf x}-{\bf X}_1| <  d$) it holds:
\begin{eqnarray}
 { \Phi}^{(E | {\rm out})}_{p'l'm'}({\bf x}-{\bf X}_2) &=&\sum_{pl}{\cal U}^{(21)}_{p'l' ; pl}(d)\,{ \Phi}^{(E | {\rm reg})}_{plm}({\bf x}-{\bf X}_1)\;,\; \\
  { \Phi}^{(E | {\rm out})}_{p'l'm'}({\bf x}-{\bf X}_1) &=&\sum_{pl}{\cal U}^{(12)}_{p'l' ; pl}(d)\,{ \Phi}^{(E | {\rm reg})}_{plm}({\bf x}-{\bf X}_2)\;, \label{transl}
\end{eqnarray}

It can be shown that the relation holds:
\be
{\cal U}^{(21)}(d)={\cal U}^{(12)\dagger}(d)\;.
\ee

By applying the operator ${\cal L}=-({\rm i}/\kappa)\,  \nabla \times$  to both members of the  Equation (\ref{transl}), and recalling the definition of the magnetic partial waves Equation (\ref{Hpw}) one finds:
\begin{eqnarray}
 { \Phi}^{(H | {\rm out})}_{p'l'm'}({\bf x}-{\bf X}_2) &=&\sum_{pl}{\cal U}^{(21)}_{p'l' ; pl}(d)\,{ \Phi}^{(H | {\rm reg})}_{plm}({\bf x}-{\bf X}_1)\;,\\
  { \Phi}^{(H | {\rm out})}_{p'l'm'}({\bf x}-{\bf X}_1) &=&\sum_{pl}{\cal U}^{(12)}_{p'l' ; pl}(d)\,{ \Phi}^{(H | {\rm reg})}_{plm}({\bf x}-{\bf X}_2)\; 
\end{eqnarray}
which shows that the translation matrices of the magnetic partial waves coincide with those for the electric partial waves.

\vspace{6pt} 



 \vspace{6pt} 

\authorcontributions{{Both authors contributed equally to the present work. Both authors have read and agreed to the published version of the manuscript. } 
}

\funding{{Not applicable. } 
}

\institutionalreview{{Not applicable.  } 
}

\informedconsent{{  Not applicable.} 
}

\dataavailability{{Not applicable.  } 
}

\conflictsofinterest{{The authors declare no conflict of interest.  } 
} 

\vspace{12pt}

\end{paracol}
\reftitle{{References} 
}

\end{document}